# Resting State Functional Connectivity Patterns Associate with Alcohol Use Disorder Characteristics: Insights from the Triple Network Model.

Daniel Guerrero[1,2], Mario Dzemidzic[3,4], Mahdi Moghaddam[1,2], Mintao Liu[1,2], Andrea Avena-Koenigsberger[4,5], Jaroslaw Harezlak[6], David A. Kareken[3], Martin H. Plawecki[7], Melissa A. Cyders[8], Joaquín Goñi[1,2,9,*]

[1] Edwardson School of Industrial Engineering, Purdue University, West-Lafayette, IN, USA
[2] Purdue Institute of Integrative Neuroscience, Purdue University, West-Lafayette, IN, USA
[3] Department of Neurology, Indiana University School of Medicine, Indiana Alcohol Research Center, Indianapolis, IN, USA
[4] Department of Radiology and Imaging Sciences, Indiana University School of Medicine, Indianapolis, IN 46202, USA
[5] Center for Neuroimaging, Department of Radiology and Imaging Sciences, Indiana University School of Medicine, Indianapolis, IN 46202, USA
[6] Department of Epidemiology and Biostatistics, Indiana University Bloomington, Bloomington, IN, USA.
[7] Department of Psychiatry, Indiana University School of Medicine, Indianapolis, IN USA
[8] Department of Psychology, Indiana University Indianapolis, Indianapolis, IN, USA.
[9] Weldon School of Biomedical Engineering, Purdue University, West-Lafayette, IN, USA
*jgonicor@purdue.edu

## ABSTRACT

Prolonged alcohol use results in neuroadaptations that mark more severe and treatment-resistant alcohol use. The goal of this study was to identify functional connectivity brain patterns underlying Alcohol Use Disorder (AUD)-related characteristics in fifty-five adults (31 female) who endorsed heavy alcohol use. We hypothesized that resting-state functional connectivity (rsFC) of the Salience (SN), Frontoparietal (FPN), and Default Mode (DMN) networks would reflect self-reported recent and lifetime alcohol use, laboratory-based alcohol seeking, urgency, and sociodemographic characteristics related to AUD. To test our hypothesis, we combined the triple network model (TNM) of psychopathology with a multivariate data-driven approach, regularized partial least squares (rPLS), to unfold concurrent functional connectivity (FC) patterns and their association with AUD-related characteristics. We observed three concurrent associations of interest: i) drinking and age-related cross communication between the SN and both the FPN and DMN; ii) family history density of AUD and urgency anticorrelations between the SN and FPN; and iii) alcohol seeking and sex-associated SN and DMN interactions. These findings provide an integrative interpretation for many individual findings reported in the literature relating functional connectivity signatures and AUD factors. Moreover, we identified a set of neural mechanisms and brain regions concomitant with AUD-related characteristics that can serve as potential treatment targets across clinical and preclinical models.

# 1. Introduction

Alcohol use disorder (AUD) is a chronic relapsing brain disorder affecting over 28 million people in the United States alone (SAMHSA, 2023). Chronic alcohol use leads to neuroadaptations that result in decreased sensitivity to alcohol, continued drinking despite negative consequences, negative reinforcement consumption, and relapse (Koob & Volkow, 2010), all of which mark more severe and treatment-resistant alcohol use. As AUD progresses, brain networks implicated in alcohol use shift away from reward to circuits underlying decision making, negative affect, memory, and craving (Koob & Volkow, 2010). In this work, we combine theory- and data-driven approaches to identify underlying neural substrates and alterations of between- and within-network interactions that characterize known AUD-related characteristics known to impart risk for AUD.

Functional balance among neural networks is necessary for adaptive cognitive and behavioral function (Friston, 2011; Wang et al., 2021). The Triple Network Model (TNM) posits that the salience network (SN) mediates switching between the default mode (DMN) and frontoparietal (FPN) networks (V. Menon, 2011; Sridharan et al., 2008) to accomplish a balance between endogenously and exogenously driven mental activity. The FPN is responsible for high-level cognitive functions and goal-oriented tasks (Miller & Cohen, 2021), and comprises the lateral prefrontal cortex, anterior inferior parietal lobule, and middle frontal gyrus (Uddin et al., 2019). The SN and FPN are active during tasks, such as those requiring external cognitive demands (Fox et al., 2005; Seeley et al., 2007; Uddin & Menon, 2009). In contrast, the DMN is most active in absence of external task demands (Greicius et al., 2002; Raichle et al., 2001), and more implicated in internal mental processes (e.g., introspection, future planning, mind wandering). The DMN includes the ventral and dorsal medial prefrontal cortex, anterior and posterior cingulate as well as retrosplenial cortex, precuneus (mostly ventral), inferior parietal lobule, lateral temporal cortex, and hippocampal formation (Buckner et al., 2008). The SN plays a central role in balancing these (and other) functional networks by detecting and prioritizing sensory input to guide attention, attending to motivationally salient stimuli, and recruiting appropriate functional networks to modulate behavior (Peters et al., 2016). Under the TNM, cognitive and emotional dysfunction from psychopathology reflects disruption in the functional integration among these three networks, both at rest and during task (Elsayed M, 2024; B. Menon, 2019).

Alcohol exposure alters neural circuits and diminishes cognitive capacity (Fritz et al., 2019; Shokri-Kojori et al., 2017; Squeglia et al., 2014; Topiwala et al., 2022; Wagner et al., 2006; Zahr & Pfefferbaum, 2017) among other negative consequences. Within- and between-alterations in SN, FPN and DMN functional connectivity patterns are particularly important in AUD (Canessa et al., 2021; Chanraud et al., 2011; Elsayed M, 2024; Maleki et al., 2022; Suk et al., 2021). With greater frequency, quantity, and duration of drinking, neuroadaptations in the brain will further alter drinking behavior itself (Koob & Volkow, 2010; Greenfield et al., 2014; Nieto et al., 2021). More specifically, regions within the DMN that are normally deactivated during task processing in healthy controls exhibit the opposite pattern in individuals with AUD, reflecting dysregulation and compromised functional connectivity (Chanraud et al., 2011; Schulte et al., 2012). Similarly, dysregulation and decreased FPN connectivity during task execution (Maleki et al., 2022; Squeglia et al., 2014), as well as slow decision-making and abnormalities in SN circuits (Galandra et al., 2018; Suk et al., 2021), are common in those with AUD.

Multiple factors contribute to the onset and progression of AUD. Impulsivity is a key risk factor for AUD (Shin et al., 2012), particularly driven by emotionally provoked rash action (i.e., "urgency"; Cyders & Smith, 2008; Whiteside & Lynam, 2001; Zorrilla & Koob, 2019). It also contributes to greater drinking quantity and frequency over time (Littlefield et al., 2014), earlier onset of AUD (Dick et al., 2010), and to a worse treatment response (Hershberger et al., 2017; Whitt et al., 2019). With greater severity, the transition to AUD is theorized to reflect the change from impulsive to compulsive alcohol use. Urgency likely reflects a shift from flexible, action-outcome decision making toward more habitual, stimulus-response behavior. This shift has been associated with disrupted interactions within and between the FPN, SN, and reward-related networks, consistent with reduced regulatory control and heightened sensitivity to salient cues (Fan et al., 2017; Giuliano et al., 2019; Gürsel et al., 2018; O'Tousa & Grahame, 2014; Su et al., 2024). Those with a biological family history of AUD are at higher risk to develop AUD (Kareken et al., 2010; NIAAA, 1997; Nurnberger et al., 2004; Oberlin et al., 2013), with evidence suggesting altered transitions between cognitive states implicating altered functional networks related to TNM (Amico et al., 2020). Finally, males and younger individuals tend to engage in heavier drinking, although negative consequences for drinking are worse in females and, recently between male and female drinking and AUD-rates have narrowed (Keyes et al., 2010). While not specific to AUD, sex and age differences have been identified in various conditions implicating the FPN, SN, and DMN and/or their within- and across network connectivity (Cummings et al., 2020; de Dieu Uwisengeyimana et al., 2022; Ernst et al., 2019; Helpman et al., 2021; Lawrence et al., 2020).

In this study we investigated the multiple associations between AUD-related characteristics and their corresponding functional neural substrates, using the TNM framework (V. Menon, 2011) to assess how the FPN, SN, and DMN and their interactions relate to factors conferring risk for AUD. The model theorizes that the SN mediates communication between FPN and DMN and that dysregulation between these three networks leads to cognitive and emotional disorders (Figure 1A). While this framework has been applied to several psychiatric and behavioral disorders, including AUD (Elsayed M, 2024), its utility in elucidating the dimensional neurobiological correlates of alcohol use and AUD remains underexplored. Recent findings (Elsayed et al., 2024) demonstrated that altered connectivity within TNM elements relates to AUD severity and recent drinking patterns in heavy drinkers. The current study extends this research by examining a broader constellation of AUD-related characteristics, including individual differences in personality traits, demographic factors known to modulate AUD risk, and an experimentally assessed index of alcohol-seeking behavior. By integrating these diverse behavioral, dispositional, and laboratory-based measures within the TNM framework, our approach offers a more comprehensive, mechanistically-grounded account of how large-scale network interactions relate to vulnerability and expression of alcohol-related behaviors.

Applying the TNM to AUD has strong scientific premise, as these networks regulate and balance relationships between introspective states (such as those related to urges and internal visceral sensation) and directed attention to either external cognitive demands (V. Menon, 2011; Sridharan et al., 2008) or alcohol-related cues and phenomena (Suk et al., 2021).

We therefore hypothesized that top-down regulating mechanisms between the SN, FPN and DMN and their intrinsic functional connectivity patterns would be disrupted reflecting AUD symptoms, alcohol use, alcohol seeking, urgency, and family history of AUD. In this study, we applied a data-driven regularized partial least squares (rPLS) approach to elucidate how resting state functional connectivity related to these characteristics and to provide fine-grained description revealing specific regions contributing to these associations. By combining the TNM with a multivariate, theory-guided analytic approach, the present study provides a novel, mechanistically grounded characterization of how large-scale brain functional networks relate to vulnerability and expression of alcohol-related behaviors, offering refined insight into potential vulnerability markers and targets for future intervention research.

## 2. Methods

## 2.1 Participants information

Participants were recruited to ensure a range of lifetime drinking history and, for safety, sufficient recent experience with alcohol and provided informed consent for all experimental procedures as approved by the Indiana University Institutional Review Board (Garrison G. & Wu, 2025). Within this larger study comprising 127 participants, a sample of 84 participants completed two alcohol infusion study sessions as described in Section 2.3. Within this sample, 55 right-handed individuals (31 female, 33 white, mean age=32.18, SD=9.9) completed an optional imaging session (see Table 1 and Section 2.3 for details). Participants were healthy, community dwelling adults who reported current heavy alcohol use (average 21.20 drinks per week (SD=26.78)). They were recruited (from the general Indianapolis community via flyers and online advertisements) to ensure both a range of lifetime drinking history and, for safety, sufficient recent experience with alcohol. Inclusion criteria included self-reported good health, aged 21-55, able to understand/complete questionnaires and procedures in English, body mass index between 18.5 and 32 kg/m$^2$. Exclusion criteria were contraindications to imaging (e.g., metal in body, claustrophobia), left-handedness, pregnancy or breast-feeding, desire to be treated for AUD or any substance use disorder or court ordered not to drink alcohol, medical/mental health conditions or medications that may influence data quality or participant safety, and evidence of substance intoxication (positive urine drug screen for amphetamines/methamphetamines, barbiturates, benzodiazepines, cocaine, opiates, or phencyclidine and associated alteration of vital signs or subjective assessment consistent with intoxication) or positive breath alcohol reading on arrival on any study day. Upon arrival at each study session, interim medical history and medication use were determined, and participants rescreened for drug use, pregnancy, and BrAC. Participants who smoked were offered nicotine replacement during their alcohol administration sessions to avoid withdrawal.

## 2.2 Participants measures

*Demographics.* Participants self-reported their age, biological sex, race, ethnicity, and highest level of completed education. Sex was coded as 1=male and 0=female. In addition, participants underwent evaluations for the following measures.

*The Semi-Structured Assessment of the Genetics of Alcoholism* (Bucholz et al., 1994). We used the alcohol module of the SSAGA to estimate a lifetime diagnosis of DSM-5 AUD (2+ symptoms). Of the 55 participants in the sample, 34 (18 women, 16 men) met the criteria for AUD (61.8%), with 21 falling in the mild, 7 in the moderate, and 6 in the severe categories. The family history module was used to quantify the presence of AUD in first- and/or second-degree relatives for each participant. A Family history density (FHD) score (Stoltenberg & Dd, 1998) was calculated for each participant, based on the degree of biological relatedness, in which parents and full-siblings with a lifetime history of DSM-5 alcohol dependence contributed 0.5 for each person, each dependent grandparent or sibling of parents contributed 0.25, and non-affected biological relatives contributed zero. We calculated FHD as the sum of weights divided by the number of counted relatives.

*Timeline Follow-back of Alcohol Us*e (TLFB; Sobell & Sobel, 1992). Recent drinking was characterized by using the TLFB procedure to provoke participant recollection of how many drinks they had on any drinking occasion over the previous thirty-five days. The following variables were calculated: number of heavy drinking days/week ($TLFB_{HDDW}$; defined as ≥5/4 drinks on an occasion for males/females respectively), number of drinking days/week ($TLFB_{DDW}$), average number of drinks/drinking day ($TLFB_{DDD}$), and the greatest number of drinks consumed on any single drinking day ($TLFB_{GDD}$). The Timeline Follow-back has been shown to produce valid measures of one's recent drinking behaviors (Sobell & Sobel, 1992).

*Concordia Lifetime Drinking Questionnaire* (Chaikelson et al., 1994). The Concordia scale is a self-report measure of total amount of alcohol drinking across the lifespan. Individuals report current and historical alcohol use, including age when alcohol use began, patterns of drinking (and changes in them), and quantity and frequency of drinking. Information is then summed to create several measurements, including the total amount of alcohol (assessed in kilograms) consumed over one's lifetime to date ($LDH_{KG}$), which is the score used in the current study. The scale was shown to provide reliable and valid information concerning one's lifetime drinking (Chaikelson et al., 1994; Koenig et al. 2009; Jacob et al. 2006; Skinner et al. 1982, Arico et al. 1995).

*The Short UPPS-P Impulsive Behavior Scale* (Cyders et al., 2014). The Short UPPS-P is a 20-item self-report scale measuring five related, although distinct, tendencies toward rash action. Only the positive and negative urgency scales were used in this study. Items are asked using a four-point Likert scale from 0 (Agree Strongly) to 4 (Disagree Strongly). Items are reverse-scored and averaged so that higher scale scores reflect greater impulsive tendencies.

*Alcohol Seeking*. Alcohol seeking was assessed using intravenous alcohol self-administration. The Computer-assisted Alcohol Infusion System software was used to compute alcohol infusion rate profiles required for exposure control (Zimmermann et al., 2008, 2009). Infusion sessions began with a priming interval, during which participants' breath alcohol concentration was increased to 60 mg/dL over 15 minutes and subsequently maintained for approximately 25 minutes to assess subjective and physiological sensitivity to alcohol. Following the priming interval, participants completed a 2.5-hour self-administration session, which included the Constant Attention Task (CAT; Plawecki et al., 2013) to earn alcohol or an alternative reinforcer (in this case, water) reward. The procedure required an escalating number of successful CAT trials to receive either of two rewards (each on independent schedule), with task difficulty adjusted to maintain ~50% success rate. Consistent with our prior work, alcohol seeking was quantified as cumulative work for alcohol or water (cwa and cww respectively), corresponding to the total number of trials each participant completed to earn alcohol or water infusion rewards (Plawecki et al., 2013, 2018). Alcohol seeking was conducted under two conditions: 1) neutral (N), and 2) aversive (A), with seeking in the presence of aversive sights and sounds (Garrison G. & Wu, 2025) modeling the transition to compulsive use.

| Participant characteristics (N=55) | | | |
|---|---|---|---|
| | Mean (SD) | Range | Units |
| Sex | 24 M, 31 F | | |
| Age | 32.18 (9.99) | 21 - 55 | Years |
| Education | 15.40 (2.08) | 11 - 20 | Years |
| Family History of AUD Density (FHD) | 0.07 (0.12) | 0 - 0.42 | Density |
| AUD Symptoms | 2.5 (2.36) | 0 - 10 | Scalar |
| Drinks per Drinking Day ($TLFB_{DDD}$) | 5.26 (4.10) | 1.7 - 26 | Standard drinks |
| Heavy Drinking Days per Week ($TLFB_{HDDW}$) | 1.72 (1.72) | 0 - 7 | Days |
| Drinking Days per Week ($TLFB_{DDW}$) | 3.80 (1.70) | 0 - 7 | Days |
| Drinks per Week ($TLFB_{DW}$) | 21.20 (26.78) | 3.2 - 182 | Standard drinks |
| Greatest Number of Drinks in a Single Day ($TLFB_{GDD}$) | 10.44 (6.45) | 3 - 32 | Standard drinks |
| Lifetime total alcohol consumption ($LDH_{KG}$) | 183.56 (348.10) | 5.3 - 2185.87 | Kilograms |
| Positive Urgency | 6.55 (2.54) | 4 - 12 | Scalar |
| Negative Urgency | 8.07 (2.64) | 4 - 15 | Scalar |
| Neutral Condition Cumulative Work for Water (N_cww) | 200.73 (203.60) | 0 - 771 | Number of completed CAT Trials |

| Neutral Condition Cumulative Work for Alcohol (N_cwa) | 254.84 (216.52) | 1 - 707 | Number of completed CAT Trials |
|---|---|---|---|
| Aversive Condition Cumulative Work for Water (A_cww) | 180.89 (173.79) | 0 - 616 | Number of completed CAT Trials |
| Aversive Condition Cumulative Work for Alcohol (A_cwa) | 270.65 (195.78) | 1 - 708 | Number of completed CAT Trials |

**Table 1.** Participant phenotypes. FHD: Biological Family History of Alcoholism (Stoltenberg & Dd, 1998). CAT: Constant Attention Task (Plawecki et al., 2013). A standard alcohol drink in the U.S. contains 0.6 fluid ounces or 14 grams of pure alcohol. This amount is equivalent to 12 ounces of regular beer (5% alcohol by volume or ABV), 5 ounces of wine (12% ABV), or 1.5 ounces of distilled spirits (80-proof or 40% ABV).

## 2.3 General Procedures

Participants completed two intravenous alcohol self-administration sessions, one in which working for alcohol was paired with aversive stimuli and another pairing work with neutral stimuli, using a progressive ratio paradigm (see full methods and results from behavioral session in Clinicaltrials.gov; Garrison G. & Wu, 2025), followed by a resting-state fMRI session. All sessions were conducted on separate days, with the fMRI session at least a week after the second infusion session (modal days = 7, median days = 13, mean days = 26.5). All study procedures occurred on weekdays, with the imaging visit scheduled in the afternoon and at least a week after participants successfully completed their second alcohol infusion session.

## 2.4 Image acquisition

Imaging was conducted on a Siemens 3T Prisma (Erlangen, Germany) MRI scanner with a 64-channel head coil array. A high-resolution anatomical volume 3D Magnetization Prepared RApid Gradient Echo sequence (MPRAGE; Lifetime Human Connectome Protocol parameters: 1 slab with a 50% distribution factor, 208 sagittal slices/slab, slice oversampling 23.1%, 0.8 mm slice thickness, 256 mm field-of-view (FoV), 93.8% FoV phase, 320×320 matrix, repetition/echo/inversion time TR/TE/TI= 2400/2.22/1000 ms, flip angle= 8 deg, GRAPPA acceleration= 2, 0.8×0.8×0.8 $mm^3$ voxels) was acquired first. Participants then completed a resting-state fMRI (rsfMRI) scan with an instruction to think about nothing in particular while fixating their gaze on a centrally located white crosshair shown on a black background once the scan began. This eight-minute blood oxygenation level dependent (BOLD) rsfMRI scan used a multiband echo-planar imaging (EPI) sequence (Center for Magnetic Resonance Research at the University of Minnesota, gradient echo, 616 BOLD volumes, TR/TE= 780/29ms, flip angle 54 deg, field-of-view 220×220 $mm^2$, matrix 88×88, fifty-five 2.5 mm thick slices, 2.5×2.5×2.5 $mm^3$ voxel, slice acceleration factor= 5) (S. M. Smith et al., 2013). BOLD fMRI acquisition was preceded by a pair of phase-reversed spin echo field mapping scans (3 A-P and 3 P-A phase direction volumes, TR/TE= 1200/64.40 ms); other imaging parameters matched the rsfMRI acquisition.

## 2.5 Image Preprocessing

Preprocessing was completed with an in-house Bash and Python 3.6 based pipeline using FMRIB Software library (FSL version 6.0.1). T1-weighted MPRAGE image of each participant was denoised prior to brain masking and extraction with ANTs (Avants et al., 2009) and then nonlinearly transformed (FSL's *flirt* and *fnirt*) to Montreal Neurological Institute (MNI) MNI152 standard space. This MNI-to-T1 transformation was followed by T1-to-EPI transformation (see EPI preprocessing) allowing standard-to-native (i.e., MNI-to-EPI) and inverse (i.e., EPI-to-MNI) transformations required to apply standard space atlases. rsfMRI data were processed in native BOLD EPI space of each participant. The preprocessing steps included BOLD volume distortion correction using FSL's *topup/applytopup* (utilizing phase-reversed spin echo field mapping scans), head motion realignment (*mcflirt*), T1-to-EPI registration (linear, nonlinear, and boundary-based registrations), normalization to mode 1000, and spatial smoothing with a 6 mm isotropic full width at half maximum Gaussian kernel.

Following recommendations for robust preprocessing (Eklund et al., 2016), the preprocessed data were entered into FSL's MELODIC (Nickerson et al., 2017) for independent components analysis (ICA)-based denoising with ICA-AROMA (Pruim, Mennes, Buitelaar, et al., 2015; Pruim, Mennes, van Rooij, et al., 2015). A single step regression was applied to the denoised BOLD volumes to avoid reintroducing artifacts in the preprocessed denoised data (Lindquist et al., 2019; Phạm et al., 2023). Specifically, regressors were applied that 1) indexed head motion using the realignment and their derivatives (Power et al., 2015), 2) accounted for physiological noise (first five signals obtained by PCA from the white matter and cerebrospinal fluid-eroded masks; an implementation of aCompCor (Muschelli et al., 2014), 3) performed high-pass filtering ($f_{min}$ = 0.009 Hz) using Discrete Cosine Transforms bases (Shirer et al., 2015), and 4) included outlier volume despiking (Phạm et al., 2023). The outliers were determined using the significant "DVARS" metrics obtained on the single-regression preprocessed data as described in (Phạm et al., 2023). This procedure tagged a mean of 1.31% (Standard Deviation: 1.24%; range: 0 – 7.29%) of residual high head motion volumes across all participants.

Individual-level rsFC matrices were determined by pairwise Pearson correlation coefficients of mean regional BOLD time-series. We implemented the Schaefer 300 cortical parcellation (Schaefer et al., 2018) and the 32-region Scale II Melbourne Subcortical Atlas (Tian et al., 2020) to assess functional connectivity among 332 brain region pairs (cerebellum excluded). To facilitate the interpretation, we aggregated the cortical regions into seventeen resting-state functional networks (RSNs) as proposed by Yeo (Thomas Yeo et al., 2011).

## 2.6 Selection and processing of participant phenotypes.

As the TLFB and lifetime drinking history variables were skewed, with long positive tails, they were logarithmically transformed (see Figure S1). The behavioral alcohol seeking variables were transformed to capture the contrast between working for alcohol and water in the neutral and aversive sessions and termed alcohol preference (neutral_ap and aversive_ap respectively; see Figure S2):

$$neutral_ap = n_cwa - n_cww$$

$$aversive_ap = a_cwa - a_cww$$

All variables were then Z-scored to standardize the magnitudes among variables presented to the subsequent learning algorithms. We then applied Principal Component Analysis (PCA) to project these AUD-related characteristics into a new set of orthogonal variables. Only PCA components/latent variables with eigenvalues greater than one were preserved and analyzed. Positive and negative urgency were compressed into a single urgency variable reflecting disposition to rash action in response to emotion, regardless of the valence (75% explained variance) and labeled “Urgency” (see Figure S3). Drinking history variables were compressed into one component that reflected a combination of recent and lifetime consumption patterns (71% explained variance) and labeled “Drinking” (see Figure S4). Alcohol seeking was further compressed into a single behavioral variable that encodes the general willingness to seek alcohol across both sessions (74% explained variance, which happened to be the mean value of neutral_ap and aversive_ap) and labeled “Alcohol seeking” (see Figure S5). In this cohort, there were no significant differences in alcohol-seeking measures by sex (unpaired t-test, $t$=1.10 $p$=0.28, Figure S6). After these preparatory steps, the final set of participant characteristics included eight representative variables: sex (male/female), age (in years), education (in years), FHD, AUD symptoms, Drinking, Urgency, and Alcohol seeking (Table 2; see Figures S7 and S8).

| **Participant Phenotypic Characteristics (N=55, female=31)** | |
|---|---|
| | Description |
| Sex | Self-reported |
| Age | Chronological age (years) |
| Education | Amount of education (years) |
| FHD | Biological family history density of alcohol use disorder |
| AUD Symptoms | AUD symptom count from the SSAGA interview |
| Drinking | 1st PCA ($TLFB_{DDD}$, $TLFB_{HDDW}$, $TLFB_{DW}$, $TLFB_{GDD}$, $LDH_{KG}$) |
| Urgency | 1st PCA (Positive Urgency, Negative Urgency) |
| Alcohol seeking | 1st PCA (Aversive Condition Alcohol Preference, Neutral Condition Alcohol Preference) |

**Table 2.** Participant characteristics used in the partial least square analysis as a phenotypic domain**.**

## 2.7 The Triple Network Model

We relied on the well-established Triple Network Model (TNM) framework and focused on the functional connectivity within and between three functional networks (SN, FPN, and DMN), including the assumption of a top-down mechanism in which the SN influences both the FPN and DMN (V. Menon, 2011). Therefore, the number of brain regions decreased from 332 to 145 and allowed us to focus on neural communication patterns within these functional networks (see Figure 1B). We adopted a 17-network solution (Thomas Yeo et al., 2011), with SN comprising the Ventral Attention Network A and B, FPN comprising Control A, B and C, while the Default Mode Network included Default A, B, C functional networks. Additionally, and guided by the TNM, we excluded direct functional interactions between FPN and DMN, allowing PLS to only model the SN-FPN, and the SN-DMN interactions.

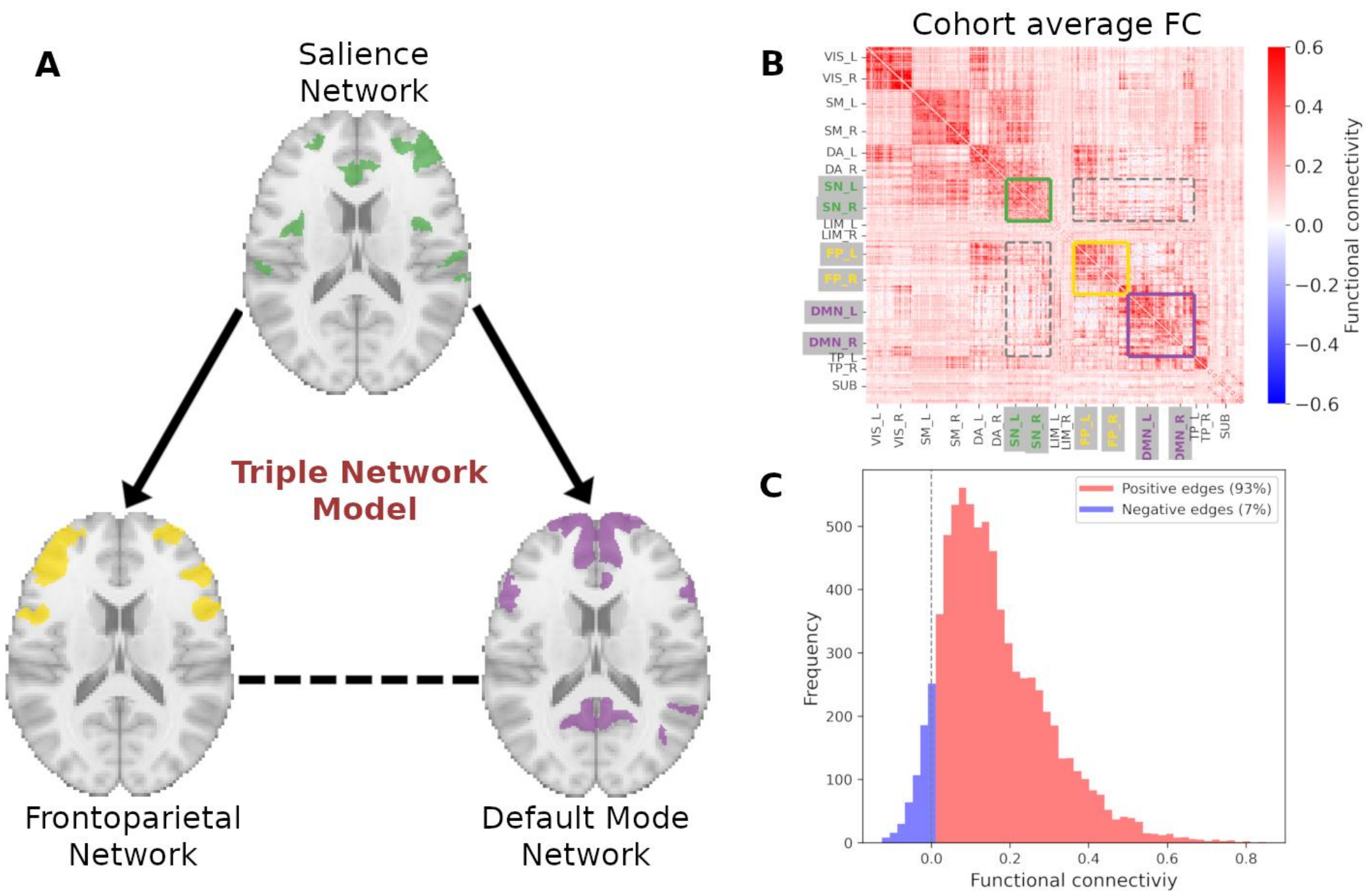

**Figure 1**. Functional connectivity within and between Frontoparietal (FPN), Default Mode (DMN), and Salience (SN) networks using a 300-region cortical parcellation (Schaefer et al., 2018) and a 17-network functional atlas (Thomas Yeo et al., 2011). A) Visual representation of the Triple Network Model (TNM), comprising SN, FPN and DMN functional networks (a representative slice shown for all three). Such model highlights the top-down communication from SN to FPN and SN to DMN (directed solid arrows) and hence, communication between FPN and DMN (dashed horizontal line) mediated by SN. B) Cohort average functional connectome highlighting functional couplings of the TNM. C) Distribution of the average functional couplings for region pairs within (full lines in B) and between the three TNM functional networks (dashed outlines in B). Vertical dashed line illustrates the division between positive and negative functional couplings.

## 2.8 Partial Least Squares

Partial Least Squares (PLS) is an unsupervised method designed to find two-dimensional components representing intrinsic relations between two sets of variables (domains). It can be considered an extension of linear regression models to handle multidimensional variables on two domains (Wold, 1966). Each PLS component identifies a hidden linear association between the domains by projecting them into a latent space. Such projection consists of linear combinations that maximize the covariance between the latent variables representing each domain. Here we refer to a PLS component as the two sets of coefficients and their corresponding latent space. A latent variable of a domain in a component is always orthogonal to all latent variables of the other domain for all other components. PLS has been successfully used in biological data analysis, where the number of variables (features) is considerably larger than the number of observations. Thus, PLS is well suited for brain connectomic analyses (Fornito et al., 2016; Sporns, 2011), where the number of brain regions and their interactions are considerably larger than the sample size, and the number of possible connections increases quadratically (Krishnan et al., 2011; McIntosh & Lobaugh, 2004).

In our study, we use PLS to uncover intrinsic relationships between the connectivity domain (rsFC within the TNM; denoted as $C$) and the phenotypic domain (demographics and AUD characteristics; denoted as $P$). For each component, PLS obtains two sets of coefficients that maximize the covariance between these domains. $W_C$ denotes connectivity domain coefficients (here presented as a square symmetric matrix with size 145x145, as determined by the number of TNM brain regions). $W_P$ denotes the phenotypic domain coefficients (for the 8 predictors: three demographic and five AUD-related; see Table 2 for details).

The statistical relationships between both domains capture underlying associations between participant phenotypes and rsFC. From a clinical perspective, the phenotypic variables with higher coefficients contribute more (are more relevant) to the discovered latent association. PLS models linear relationships between the $C$ and $P$ domains and assumes that they can be linearly decomposed as:

$$C = \Phi W_c + E_c$$
$$P = \Psi W_p + E_p$$

Where $\Phi$ and $\Psi$ are latent vectors representing the intrinsic association between $C$ and $P$, $W_c$ and $W_p$ are the abovementioned sets of coefficients, with $E_c$ and $E_p$ as error terms or residuals.

$P$ and $C$ coefficients are set as to maximize the covariance between the latent vectors $\Phi$ and $\Psi$ (see Figure 2).

$$\max_{\{W_c, W_p\}} \quad COV(\Phi, \Psi)$$

$$\Phi = CW_c^{-1}$$

$$\Psi = PW_p^{-1}$$

Correlated variables are common in both neuroimaging and phenotypic data. In the Partial Least Squares method, moderate correlations among variables within a domain are not problematic. For each component, the coefficients ($W_c$ and $W_p$) are estimated to distribute weights across correlated predictors in proportion to their shared covariance with the other domain, resulting in stable and interpretable latent components rather than unstable individual coefficients (Wold, 1985; Abdi, 2010). Unlike multiple linear regression, where multicollinearity can inflate variance and destabilize coefficients, PLS compresses correlated predictors into orthogonal latent directions that capture their shared information and maximize covariance with the outcome domain (Abdi & Williams, 2013; Geladi & Kowalski, 1986). Consequently, while moderate correlations (e.g., $|r| \approx 0.3$–$0.5$) may influence the relative weighting of variables, they do not compromise the stability or interpretability of the latent variables, nor the cross-domain associations extracted by PLS.

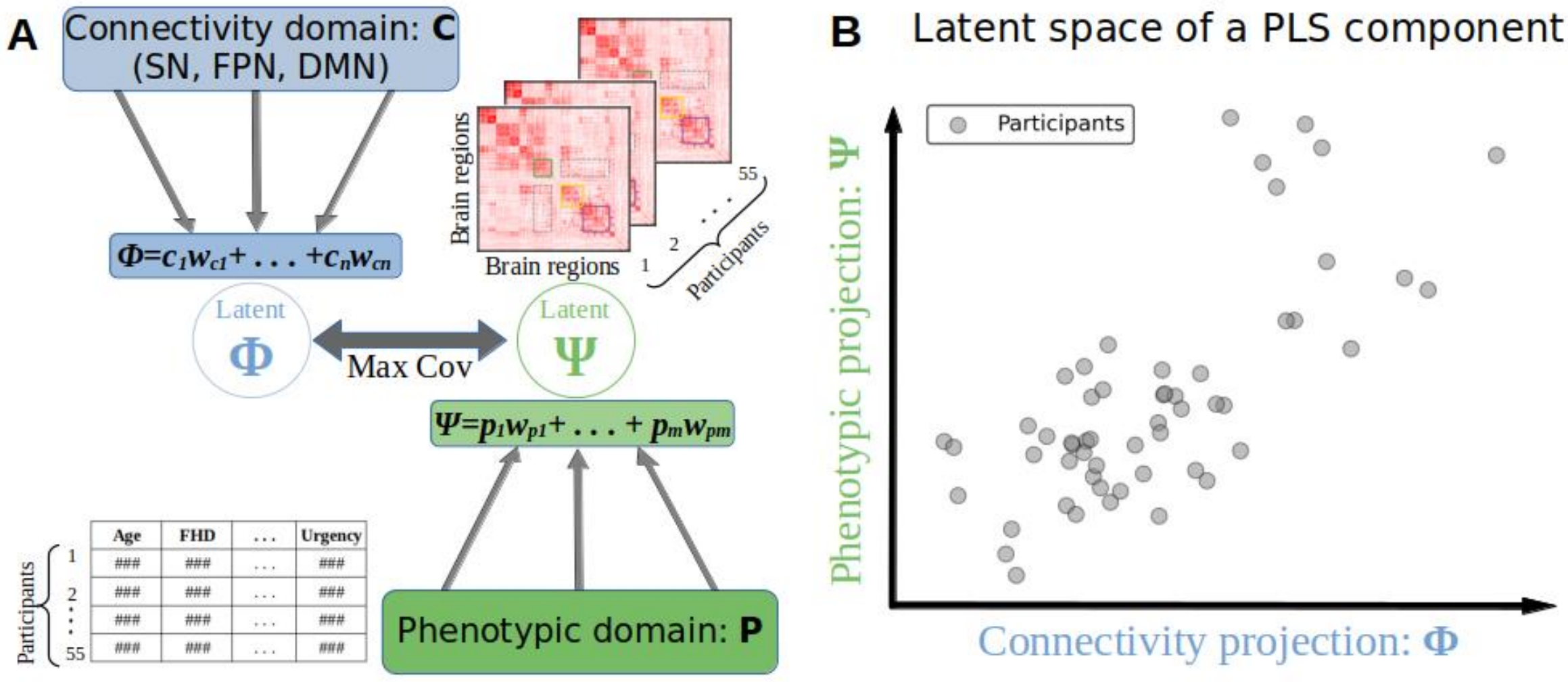

**Figure 2**. A) Schematic representation of the Partial Least Squares (PLS) analysis. Two sets of variables (here connectivity and phenotypic domains) are projected to a latent space of lower dimensionality (X and Y, respectively). For each PLS component, a linear regression is performed in the new space and the linear combinations are optimized to maximize the covariance between the variables in the latent space. B) Example of a PLS component maximizing covariance between the two domains (axes are the corresponding latent variables).

## 2.8 Regularized Partial Least Squares (rPLS)

As with many other learning methods (Tian & Zhang, 2022), regularization can also be applied in PLS. Here, we simultaneously regularized both domains to shrink their coefficients concurrently. PLS regularization is a data-driven feature subset selection strategy that adds reliability by preventing overfitting and contributes to the interpretability of the results (Tibshirani, 1996). The connectivity domain comprises brain region pairwise coupling information and is represented by high dimensional vectors. As learning methods are ill-posed when presented with high-dimensional data (i.e., more variables than samples) (Trunk, 1979), regularization addresses high dimensionality and provides fine-grained results (increased coefficient specificity) that eases interpretability.

Our goal was to identify functional couplings that play an important role in the overall relationship between the phenotypic characteristics and brain connectivity patterns. We computed both standard (i.e. non-regularized) and regularized PLS (rPLS) solutions using the Matlab implementation described in (Monteiro et al., 2016). This rPLS implementation provides flexibility for different regularization levels for each data domain by using two independent regularization parameters (one per domain). The range of possible values for each regularization parameter, $\lambda_D$, is $1 \leq \lambda_D \leq |D|^{1/2}$ where $|D|$ represents the number of variables in the respective domain (maximal regularization attained when $\lambda_D=1$). For the connectivity domain, we opted for a regularization level that preserves 50% of the original (non-sparse) solution to gain specificity on the connectivity domain. For the phenotypic domain, we selected the smallest regularization that restricts the first component to a maximum of 3 features (see Figures S9 and S10). The orthogonality between components is achieved via a deflation procedure in which new components are iteratively computed on the residuals of the previous ones (Monteiro et al., 2016).

## 2.9 Network interaction significance testing

We identified significant functional edges by applying a null model to assess the significance of each network interaction provided by the rPLS (connectivity coefficients). For each component, we constructed a null model by randomly shuffling its connectivity coefficients and compared this randomized connectivity profile with the true coefficients (see Contreras et al., 2017). Network interactions with total contribution (sum of absolute value of coefficients) higher than the corresponding null model ensemble (99th percentile; 1,000 null model runs) were considered significant. This approach allowed us to identify high-level connectivity patterns significantly represented in the connectivity domain coefficients of each rPLS component (see Figure S11).

### 2.10 Within cohort stability of rPLS components

Leave-one-out cross validation (LOOCV) assessed stability by leaving out one sample of the data at a time and training the model on the remaining samples. This process was repeated for each sample in the dataset. The model's stability is evaluated using the ensemble of resulting models and subsequent outputs (coefficients), which provides a more robust description of its behavior (Efron, 1979; Stone, 1977). We computed 55 leave-one-out PLS iterations, equal to the number of participants in the sample. Our LOOCV analyses included the coefficients distribution for each phenotypic variable, as well as the coefficients standard deviation for each functional coupling in the connectivity domain.

## 3. Results

### 3.1. Associations between AUD-related characteristics and resting-state functional connectivity

We used PLS to uncover associations of the eight phenotypic variables and rsFC patterns within *a priori* functional networks (FPN, SN, and DMN) and network interactions modeled by the TNM (Table 2). We assessed the goodness of fit for each component based on the covariance score and subsequently on the percentage of covariance with respect to the maximum (occurring, by definition, in the first component; see Figure S12). For each orthogonal component, PLS produces two coefficients sets, one for each domain (that is, AUD-related characteristics and connectivity). Each pair set describes a multivariate association between the domains and a different overall association. Figure 3 summarizes the associations between the two domains and the respective coefficients for each of the four analyzed components.

The connectivity set of coefficients obtained with PLS involves the entire connectivity domain of the TNM (that is, all edges potentially participate in the association). To uncover the most relevant functional couplings participating in the association, we used a regularized version of PLS (rPLS: Monteiro et al., 2016) as detailed in the Methods section. This resulted in regularization parameters $\lambda_C$ =49 and $\lambda_P$ =1.5 for the connectivity and phenotypic domains, respectively, for all PLS components (see Figures S9 and S10).

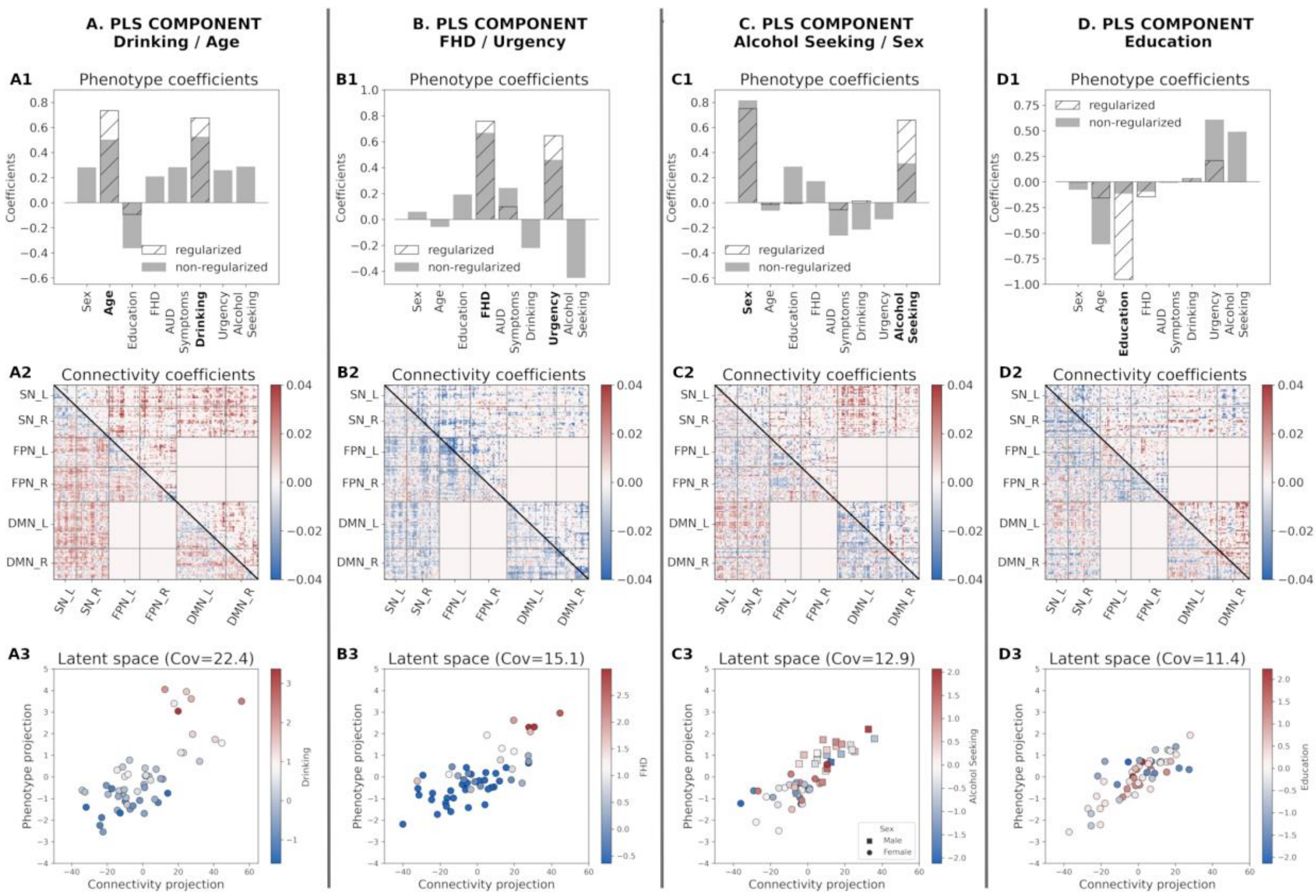

**Figure 3.** First four PLS components (non-regularized; gray bars, regularized ($\lambda_C$ =49 and $\lambda_P$ =1.5); hatched bars in the phenotype coefficients). Based on the TNM assumptions, we excluded the interactions between FPN and DMN (excluded blocks have all zeros in the Connectivity coefficient matrices). The analytically identified components were named based on their phenotypic contributors: A) ***Drinking/Age*** B) ***Family History/Urgency***, C) ***Alcohol Seeking/Sex*** and D) ***Education*** components. Bottom row shows the latent associations between the Connectivity and Phenotype domains as identified by PLS with the colored gradient representing the most prominent phenotypic variables in each component.

The results are illustrated by Figure 3. Each PLS component is named by its predominant phenotypic characteristics indicated by the magnitude of their coefficients (see Figure 3 A1-D1).

i) The *Drinking/Age* component denotes that older participants with high drinking (first principal component of recent and lifetime drinking variables) have an increased between-network interactions of the SN with both FPN and DMN and decreased within network connectivity for all three networks.

ii) The *FHD/Urgency* component denotes that participants with high family history density and high urgency have a decreased communication between the SN and FPN, as well as decreased within-network interactions in all three functional networks.

iii) The *Alcohol Seeking/Sex* component indicates that males with high alcohol seeking behavior (PCA of willingness to work for alcohol across sessions) have an increased cross communication between the SN and DMN and decreased communication within the DMN.

iv) Lastly, the *Education* component, largely driven by years of education, does not involve any of our AUD-associated traits. Hence, the subsequent analyses focus on the first three rPLS components.

## 3.2 Stability analysis of the rPLS components.

Variability across iterations was evaluated for each component and coefficient profile. LOOCV results (55 runs) show that the rPLS solution pattern derived from all participants is preserved across iterations (Figure S13). Specifically, the average effect of the dominant phenotypic variables (box plots) is well above zero and centered around the estimated coefficient values in the regularized solution for the full cohort. The standard deviation of the connectivity coefficients is small for all components, where the Drinking/Age component shows the highest stability.

## 3.3 Neural substrates of the PLS components

We determined positive and negative strengths for each region to investigate their contributions in each PLS component. Briefly, given a connectivity matrix (here a connectivity coefficients matrix), the positive strength of a region is the sum of all positive coefficients in its row, whereas negative strength is the sum of all the row's negative coefficients (Fornito et al., 2016). In our case, strength was obtained for each component similar to other functional connectivity decomposition methods (Amico et al., 2017). Positive strength summarizes the direct association between an entire region's connectivity profile and the phenotypic domain, whereas a negative strength summarizes the corresponding inverse association.

Next, we organized the signed strength profile comprising all 145 regions by functional network membership and identified the top contributing regions (top 5%, resulting in 15 regions per component). In the Drinking/Age component, the strength profile shows primarily regions with increased connectivity, mainly within the SN and the FPN (Figure 4C). The *FHD/Urgency* component profile is marked by decreased functional connectivity in the SN and the FPN (Figure 5C). The strength profile for the *Alcohol Seeking/Sex* component includes regions with increased connectivity distributed between the SN and the DMN (Figures 6C).

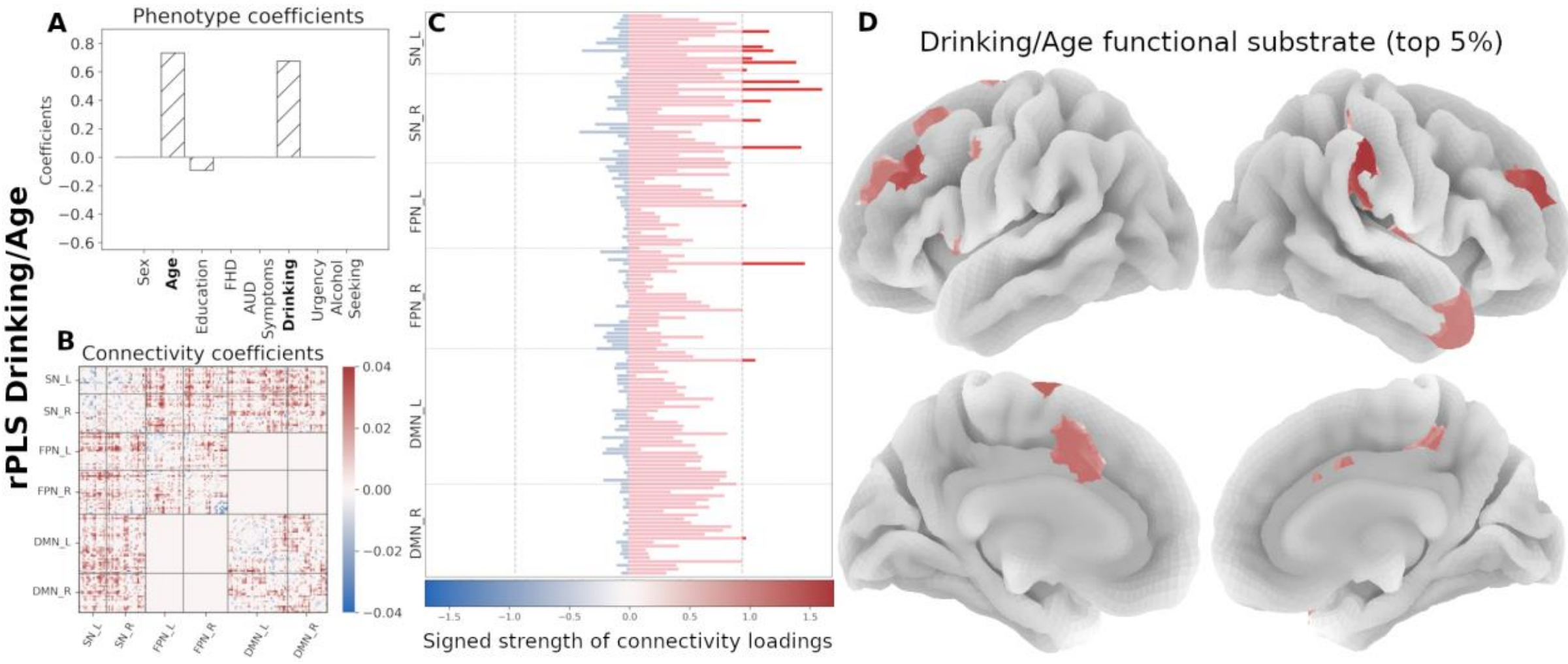

**Figure 4.** Neural substrate of ***Drinking/Age***. (A) Phenotype coefficients show Drinking and Age as the dominant and positively contributing factors. (B) Connectivity coefficients show mainly increased between-network interactions. (C) Signed strength of connectivity coefficients with the contribution for each region. The gray line indicates the 95-percentile of the strength distribution. (D) Brain rendering of the functional substrate for the top 5% regions with highest signed strength (regions above the threshold in panel C). Color gradient indicates the relative strength per region.

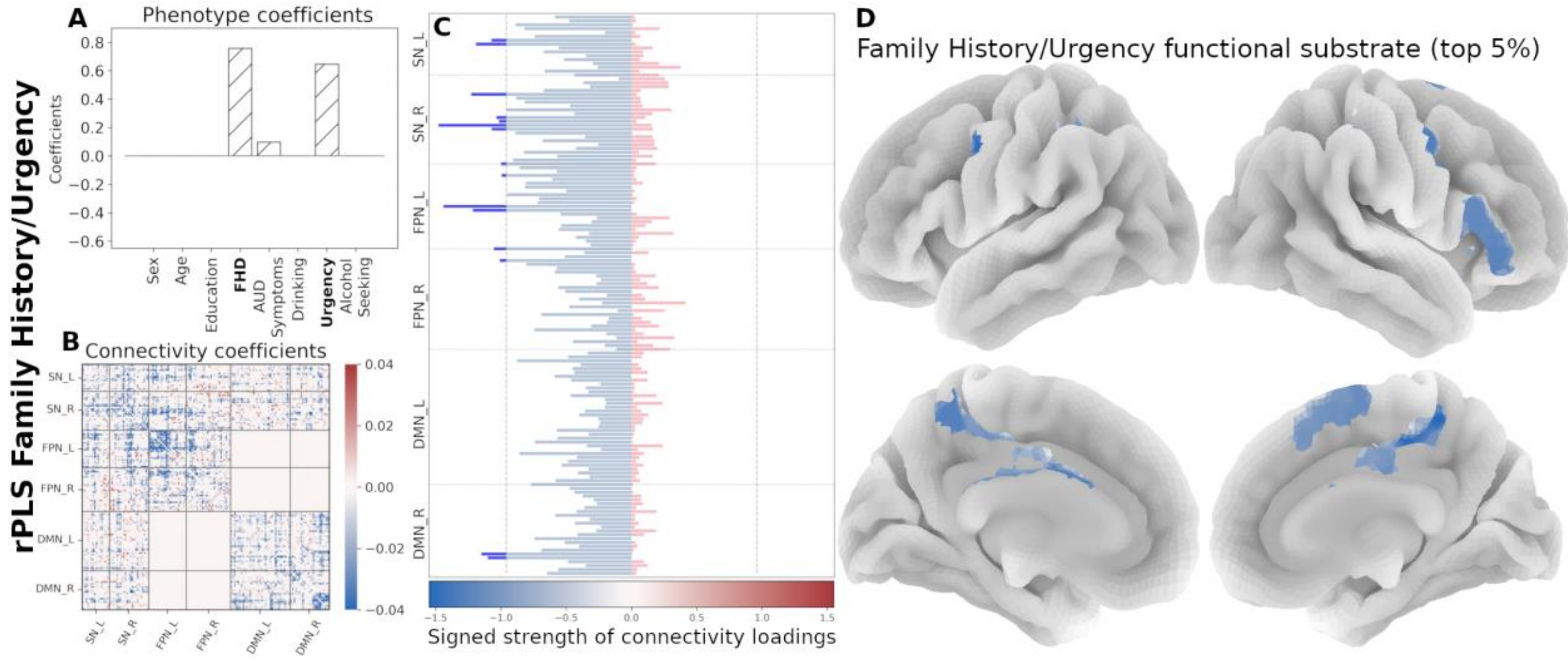

**Figure 5.** Neural substrate of *FHD/Urgency*. (A) Phenotype coefficients are dominated by FHD and Urgency, both with positive contributions. (B) Connectivity coefficients are marked by decreased interactions. (C) Signed strength of connectivity coefficients with the contribution for each region. The gray line indicates the 95-percentile of the strength distribution. (D) Brain rendering of the functional substrate for the top 5% regions with highest signed strength (regions above the threshold in panel C). Color gradient indicates the relative strength per region.

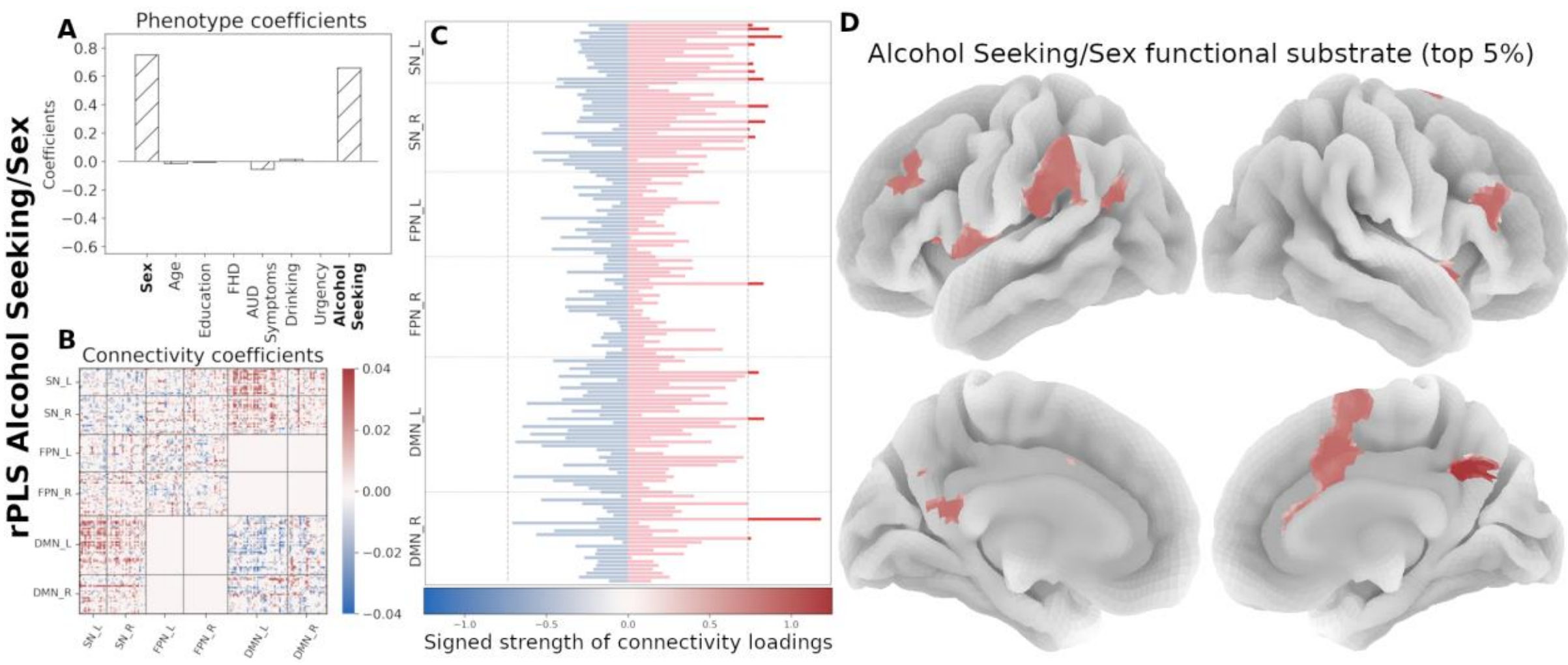

**Figure 6.** Neural substrate of *Alcohol Seeking/Sex*. (A) Phenotype coefficients are dominated by alcohol seeking (willingness to work for alcohol) and age, both positive. (B) Connectivity coefficients in this component show increased contributions. (C) Signed strength of connectivity coefficients with the contribution for each region. The gray line indicates the 95-percentile of the strength distribution. (D) Brain rendering of the functional substrate displaying the top 5% regions with highest signed strength (regions above the threshold in panel C). Color gradient indicates the relative strength per region.

In the normative TNM-based canonical resting-state connectivity circuit the SN suppresses the FPN while enhancing activity in the DMN (Figure 7A). Using this canonical TNM circuit as a reference, we summarize the results of each PLS component schematically by representing the main connectivity effects involving the three *a priori* functional networks (see Methods section and Figure S13). These diagrams capture distinct aspects of the relation between functional connectivity and AUD-related characteristics that represent high-level concurrent communication mechanisms involving SN, FPN and DMN (Figure 7). Here, the sign of the interactions is inferred from the rPLS results whereas the directionality is imposed by the TNM model.

The *Drinking/Age* circuit shows an increased communication of the SN with both FPN and DMN which, in combination with the TNM model, indicates a top-down regulation mechanism mediated by the SN. The left hemisphere of SN communicates with the left hemisphere of the FPN and the left hemisphere of the DMN while the right SN communicates with the left hemisphere of both FPN and DMN. The *FHD/Urgency* circuit is marked by decreased associations of the SN and the left FPN, as well as decreased communication within the right SN, left FPN and right DMN. Notably, communication between SN and DMN does not feature. The *Alcohol Seeking/Sex* circuit shows again findings consistent with a top-down structure, with increased communication between SN and the left DMN, while the communication within left DMN is decreased. PLS components are marked by specific connectivity patterns involving different subsets of brain regions; however, region contributions are not exclusive to a single component. Noteworthy is that components associated with drinking behavior (*Drinking/Age* and *Alcohol Seeking/Sex)* share the left hemisphere SN regions, most prominently within the prefrontal and insulo-opercular areas (overlapping regions are shown in Figure S14 and listed in Table S1).

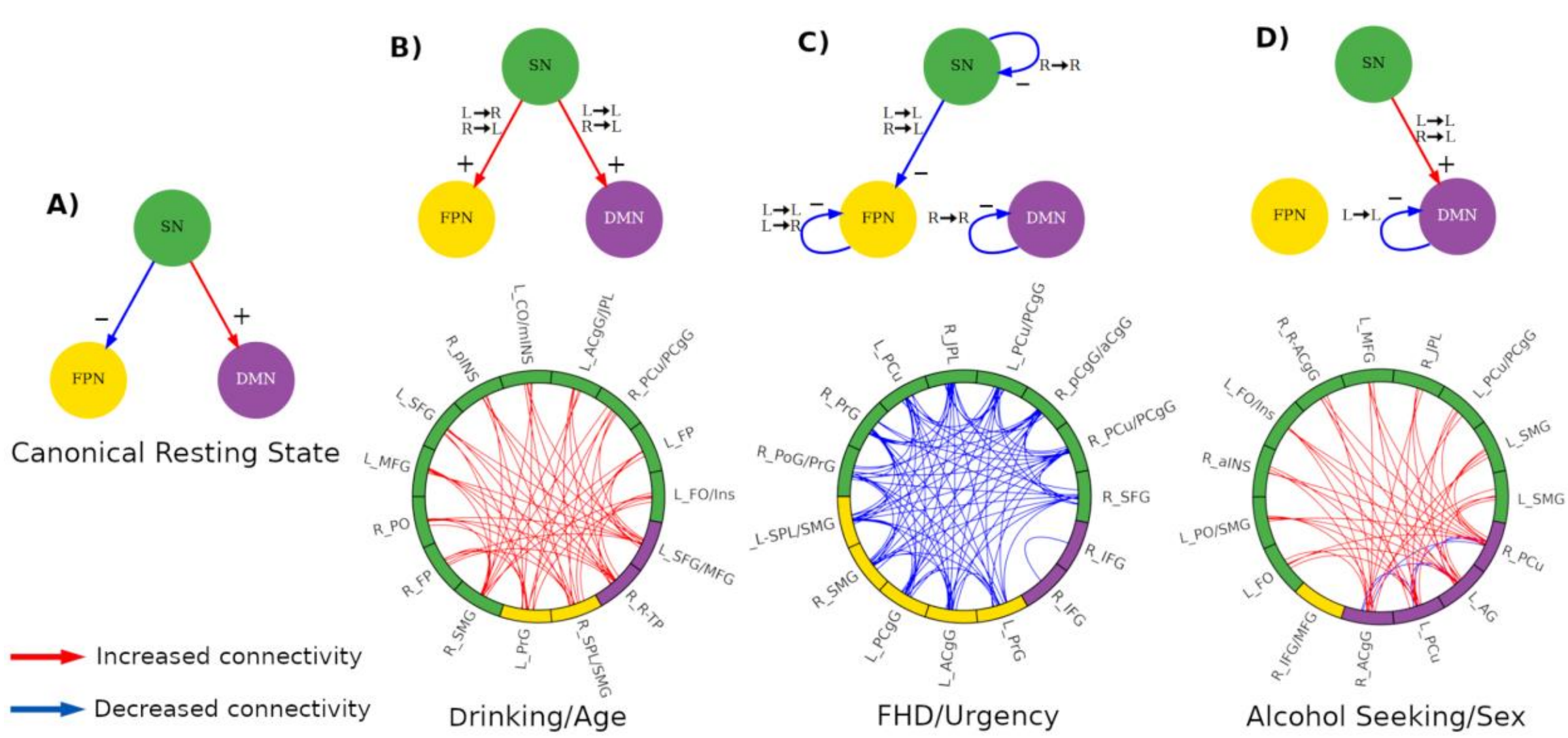

**Figure 7.** Schematic representation of the neural circuits underlying each component under the TNM and their association with AUD-related characteristics. **(A)** Canonical resting-state brain communication proposed by the TNM. **(B)** Drinking and Age are associated with an increased communication from SN to both FPN and DMN. **(C)** FHD and Urgency are associated with a decreased connectivity from SN to FPN and decreased within-network connectivity in both SN and FPN. **(D)** Alcohol seeking and Sex (being male) is associated with an increased connectivity from SN to DMN and decreased within-DMN connectivity. Abbreviations: salience network (SN), frontoparietal network (FPN), default mode network (DMN), left hemisphere (L), right hemisphere (R). (B-D) Note that the sign (color) of the network interactions is inferred from the rPLS results. All interactions follow the network directionality assumed by the TNM.

# 4. Discussion

In this work, we characterized the neural substrates and communication patterns associated with AUD-related characteristics at network and regional levels under a TNM framework, which hypothesizes that the SN mediates and recruits neural resources from the FPN and the DMN (V. Menon, 2011; Sridharan et al., 2008). Alterations in this three-network circuit appear in psychopathological conditions (Elsayed M, 2024; B. Menon, 2019; V. Menon, 2011), but the specific communication changes remain unclear. By combining the TNM, a data-driven analytical approach, resting state fMRI, and experimental design incorporating intravenous alcohol seeking paradigm in a sample of heavy drinkers including those with an AUD, we uncovered specific concurrent communication mechanisms between the FPN, SN, and DMN resting-state functional networks as they relate to different AUD-associated characteristics spanning drinking, family history of AUD, AUD symptom count, urgency, biological sex, age, and alcohol self-administration. Three major groups of AUD-related characteristics were analytically identified to be associated with connectivity patterns in the TNM circuit: a Drinking/Age component, an FHD/Urgency component, and an Alcohol Seeking/Sex component. These findings reflect the complexity and heterogeneity of AUD, with different phenotypic features associated with different interactions of the TNM functional networks. It is noteworthy that AUD symptom count (i.e. disorder severity) was not a prominent factor in any of the rPLS components. Our analysis extends previous work showing AUD-related alterations in predefined regions as a function of AUD (Elsayed M, 2024; Suk et al., 2021) by including inter-relationships between networks within the TNM framework, and as a function of AUD-related characteristics and risk factors.

The SN, FPN and DMN are involved in a wide variety of cognitive and emotional processes and their coordination is thought to properly allocate neural resources in response to endogenous and exogenous demands. In turn, a range of psychiatric disorders are thought to affect these networks (Boehm et al., 2014; Rai et al., 2021; Stern et al., 2012; Suk et al., 2021). Findings of altered functional connectivity between these three networks in AUD and risky drinking include increased connectivity between the FPN and DMN (Suk et al., 2021), high resting-state connectivity in the FPN (Sousa et al., 2019), and abnormalities in the SN (Galandra et al., 2018; Suk et al., 2021).

The *Drinking/Age* component, is marked by increased communication between the SN and both the FPN and the DMN (see Figure 7), suggesting potentially increased top-down control. Aging positively contributes to this association, as well as drinking, which reflects both recent and lifetime drinking consumption patterns (Table 2). The observation that these two variables track together may be indicative of persistent and ongoing drinking patterns in this sample of heavy drinkers. This component suggests that the SN affects both the FPN and the DMN, which is consistent with previous findings (Suk et al., 2021) that the FPN and DMN are simultaneously active during resting state. In healthy individuals at rest, the FPN and the DMN are negatively correlated, suggesting functional specificity and a mechanism to coordinate neural resources in response to internal and external cognitive demands (Deming et al., 2023; V. Menon, 2011). Positive correlation between the FPN and DMN during rest, particularly when accompanied by increased SN-FPN and SN-DMN communication, may reflect a disruption of the typical TNM architecture in which the SN flexibly coordinates engagement of these two networks. Such increased co-activation suggests reduced functional independence of the FPN and DMN, suboptimal network segregation, and a possible reliance on compensatory mechanisms to maintain cognitive or behavioral control. In this context, the observed pattern may represent an over-engagement or distortion of top-down regulatory processes, potentially reflecting the combined influence of aging and alcohol involvement on large-scale network organization. Examples of these abnormal brain configurations and their association with multiple psychopathological disorders are reported in numerous studies (Boehm et al., 2014; He et al., 2021; Rai et al., 2021; Stern et al., 2012; Suk et al., 2021; Zhang & Volkow, 2019).

The *FHD/Urgency* component was marked by decreased communication between the SN and the FPN and decreased within-network communication in all three networks. Decreased functional interactions are a marker of disassociation between brain processes, a signature of automatic processes that do not require cognitive effort (Finc et al., 2017; Kitzbichler et al., 2011). For urgency, this may be indicative of a neural configuration suited for automatic (non-regulated) predisposition towards impulsive behavior, as urgency is theorized to reflect lower top-down and higher bottom-up processing (Cyders & Smith, 2008; G. T. Smith & Cyders, 2016). Several studies have largely supported dysregulation between these two systems as related to urgency (see Um et al., 2019; Zorrilla & Koob, 2019). In addition, this component is characterized by decreased within-network connectivity in all three networks, which has been linked to impaired executive dysfunction, emotion regulation, and risk assessment (Clark et al., 2008; Grodin et al., 2017; Singer et al., 2009). Explanations for reduced connectivity within the FPN include weak intrinsic connectivity within its nodes and constrained access to salience stimuli from the SN (Weiland et al., 2014).

Recent evidence suggests that SN dysfunction affects cognitive performance in individuals with AUD, but this impairment is largely mediated through the FPN (Rawls et al., 2021). SN dysfunction also causes impaired mapping of salient events and a disrupted balance of appropriate neural resources (V. Menon, 2011), consequences of which include weak emotion regulation and a lack of cognitive control (V. Menon, 2011; Seeley et al., 2007; Zilverstand et al., 2018). Reduced connectivity within the SN has been observed in individuals after periods of acute alcohol consumption (Gorka et al., 2018), is linked to inability to restraint subjective urges (Sullivan et al., 2013) and is a predictor for future relapse (Camchong et al., 2022; Kohno et al., 2017). Our findings contrast with one resting state study that found increased within-network connectivity in the SN, orbitofrontal cortex, and the DMN and increased between-network connectivity as a function of negative urgency in those with AUD (Zhu et al., 2017). In our work, the neural circuit related to urgency also reflects family history density, suggesting that automatic engagement in rash behavior may be also related to genetic risk factors for AUD, as shown in prior studies (Dick et al., 2010; Stephenson et al., 2023).

The *Alcohol Seeking/Sex* component was characterized by increased communication between the SN and DMN and reduced communication within the left DMN. Communication between the SN and the DMN is enhanced during the withdrawal phases in addiction (Zhang & Volkow, 2019). Similarly, Arienzo and colleagues (Arienzo et al. 2020) showed increased connectivity between the salience and reward networks in young adult binge drinkers relative to social drinkers, with the salience network seed represented by the anterior cingulate cortex, which prominently contributes to the alcohol seeking component. Altered connectivity between the DMN and cortical regions associated with memory and emotion regulation is critical for compulsive drug seeking despite adverse consequences (Zhang & Volkow, 2019). Decreased connectivity within the DMN has been associated with several substance use disorders, (Vergara et al., 2017) including AUD (Müller-Oehring et al., 2015), which is reflective of reduced self-awareness and rumination during alcohol abstinence. In a recent study (Muller et al., 2024) assessing network configurations during active alcohol approach, the DMN was found to be an important network for information integration, suggesting a possible configural state that facilitates greater intensity of alcohol seeking. Sex is an important and well documented AUD risk factor (Becker et al., 2012; Flores-Bonilla & Richardson, 2020; George B. Richardson & Brian B. Boutwell, 2020; Plawecki et al., 2018). Its presence in the component could represent effects from sex alone or in combination with appetitive effort for alcohol.

Combining participant phenotypes in a clinically meaningful manner helps to understand how gradients of AUD risk relate to the TNM circuit. Each TNM circuit (Figure 7) characterizes the neural substrate representative of AUD risk factors. By comparing these with the "canonical TNM circuit" (for healthy controls in resting-state; Figure 7A), we begin to interpret how large-scale mechanisms might form signatures of these factors. For example, while both FHD and drinking are related in the population, the data here suggest that these factors involve very different signatures of communication within the TNM circuit. Data such as these may therefore facilitate targeted interventions aimed at specific clinical features.

In contrast to region specific seed-based analyses or testing univariate associations with a single phenotype, we assessed whether the communication patterns within and between TNM networks are associated with AUD-related characteristics. Using a combination of experimental design, theoretical model, and data-driven approaches is a core strength of this work. This methodology allows us to understand how complex interactions involving multiple AUD-related characteristics shape the communication patterns between *a priori* functional networks. In addition, this work extends the current application of the TNM to a sample with heavy alcohol use and provides an interpretative framework approach to better understand TNM alterations specific to different psychopathological disorders.

Critically, the LOOCV results show that the identified relationships are stable (the pattern is preserved across iterations), and the contributions (coefficients) of individual variables and brain connectivity patterns are preserved overall (Figure S13). The variables with the greatest contributions are on average the same, which means that the associations captured by the components are stable within the cohort and less likely to be driven by specific participants.

Beyond incorporating a heterogeneous combination of AUD-related traits, demographic factors, and behavioral measures, the primary contribution of our work is in summarizing these associations into a set of interpretable, top-down mechanisms that identify the most prominent regions and communication patterns within the FPN, SN, and DMN that characterize our sample of heavy drinkers. Each mechanism reflects a distinct configuration of phenotypic characteristics, helping to disentangle the multifaceted nature of AUD. While these findings do not imply immediate clinical application, they do provide a mechanistic structure that can guide hypothesis generation for future studies rather than implementing individual phenotypic correlates.

Our findings extend previous literature studying associations between resting state connectivity and AUD-related factors. Prior resting-state studies have shown that alcohol use and AUD are associated with altered connectivity in reward, salience, and control networks (Zhu et al., 2017; Vergara et al., 2017; Arienzo et al., 2020), and recent work within the Triple Network framework has linked FPN connectivity to AUD symptoms (Elsayed et al., 2024). Our results align with this literature in demonstrating large-scale network dysregulation but extend it by identifying multivariate, hierarchical connectivity components that dissociate distinct AUD-related phenotypes—including age-related drinking, family-history/urgency dimensions, and laboratory-assessed alcohol seeking—each reflected in specific SN–FPN or SN–DMN interactions. In contrast to studies emphasizing global or dynamic connectivity disruptions (e.g., Vergara et al., 2017; Zhang et al., 2024), our findings reveal trait- and behavior-specific patterns with a greater potential for finer mechanistic specificity. This approach directly addresses calls for integrative, theory-driven analyses of AUD neurobiology (Xie et al., 2022; Suk et al., 2021) and refines prior SN-focused findings (Patel et al., 2025) by pinpointing the nodes and cross-network interactions that are most strongly associated with diverse AUD-related characteristics. Due to the nature of the PLS solution, the components describing latent associations follow a hierarchical (and orthogonal) order where the first component (Drinking/Age) captures the main features of aberrant connectivity and their corresponding phenotypic signatures. Consequently, the other components (FHD/Urgency, Alcohol Seeking/Sex) describe less prominent, but more trait-specific associations well-tailored for interventional targets.

The interpretation of our results must also be considered in light of our inclusion/exclusion criteria. Other substance dependence can be comorbid with AUD and is known to alter executive function, memory, attention, and impulse control, as well as FC in relevant brain networks. Although our assessment excluded severe current substance use disorders, our results may be impacted by ongoing low-severity or past non-alcohol substance use disorders. There are other limitations. First, the modest sample size may affect the findings' replicability. Likewise, our sample is, by design, restricted to participants who endorse heavy alcohol use, with about 60% meeting criteria for AUD, which may impact the generalizability of our findings. That said, AUD symptom count was unrelated to the connectivity patterns. Second, cross-sectional data and the statistical method (PLS) preclude causal interpretations of the inferred associations and interactions between networks. Finally, the analysis of resting-state data was not complemented by the task fMRI assessments that could target specific AUD-relevant brain regions (e.g., reward system) and behaviors (e.g., alcohol cue-response, working for alcohol reward, etc.).

AUD is a clinically heterogeneous condition, and different combinations of risk factors, traits, and behavioral tendencies are likely supported by distinct patterns of large-scale network functioning. The present findings extend previous efforts (Elsayed et al., 2024) by demonstrating that multiple AUD-related characteristics—including trait-level factors, demographic features, and laboratory-based alcohol seeking—map onto specific interactions among regions within the FPN, SN, and DMN. Importantly, these associations were detected using a data-driven modeling approach nested within a well-established theoretical framework, allowing us to identify convergent network features across diverse behavioral and dispositional dimensions. The long-term goal of this line of work is to deepen mechanistic understanding of how these network interactions support vulnerability to, or protection from, AUD and ultimately to inform the development and evaluation of novel AUD interventions. Such a goal requires converging evidence across preclinical investigations that independently identify similar circuits and behavioral associations. The present results should be viewed as one step in a cumulative scientific process, providing candidate neural pathways that can be tested, refined, or challenged in animal models, where causal manipulation of homologous circuits can clarify whether the identified interactions represent viable targets for later human intervention studies. The current study therefore offers a mechanistic foundation and a set of testable hypotheses to support longer-term translational progress in AUD research. For example, while neuromodulation approaches have been applied to AUD, (Alba-Ferrara et al., 2014; Mehta & Parasuraman, 2013, Sorensen et al. 2020; Mehta et al. 2024), few targets have been explored and there is large heterogeneity in effects across individuals and technique suggesting more work is needed to determine the most effective targets as well as neuromodulatory approach. Most assuredly, there is no single brain target, but rather a more likely goal is to affect the function of broader networks. In that regard, multiple regions, such as precuneus/posterior cingulate gyrus, left precentral gyrus, and right juxtapositional lobule (see Figure S14) show a potential as novel target entry points for neuromodulation. These brain areas span core components of the TNM and participate in interoception, salience evaluation, internally directed processing, and action tendences, that map onto theoretical mechanisms underlying alcohol seeking and AUD. The recurrence of these regions across distinct behavioral and dispositional measures suggests they may represent points of convergence within the broader circuitry supporting vulnerability to AUD and can be used to refine mechanistic hypotheses that can be evaluated in future human and animal studies.

# 5. Conclusion

We combined theory- and data-driven approaches to document underlying neural substrates characterizing drinking, family history of AUD, AUD symptom count, urgency, and alcohol seeking, all factors known to impart risk for AUD. Focusing on the TNM, this study provides an approach for a comprehensive characterization of the neural components underlying AUD, revealing how the brain networks unfold into concurrent characteristic-specific configurations. This study demonstrates the utility of data-driven approaches in uncovering associations between resting-state functional substrates and phenotypic characteristics that could aid in the identification, development, and testing of novel treatment targets across preclinical and clinical models.

**Acknowledgments.** This work was supported NIH CTSI CTR EPAR2169, NIH R21 AA029614, NIH R01 AA029607, and Indiana Alcohol Research Center P60AA07611, NINDS R01NS126449, NINDS 622R01NS112303, and by Lilly Endowment, Inc., through its support for the Indiana University Pervasive Technology Institute and by Shared University Research grants from IBM, Inc., to Indiana University. We would like to thank faculty and staff in the Research Imaging Core of the Indiana University Medical Imaging Research Institute, Dr. Yu-Chien Wu and Dr. Qiuting Wen for assistance with MRI sequence development and testing as well as MRI technologists Traci Day, Robert Bryant, and Will Korst for their invaluable help during imaging. The cooperation and support of the IU Health Investigational Drug Services in the preparation of 6% alcohol v/v infusate was essential for this research project. The programming expertise of Victor Vitvitskiy was vital to this project and most sincerely appreciated. This project was funded with support from the Indiana Clinical and Translational Sciences Institute which is funded in part by Award Number UM1TR004402 from the National

Institutes of Health, National Center for Advancing Translational Sciences, Clinical and Translational Sciences Award.

# Supplementary Material

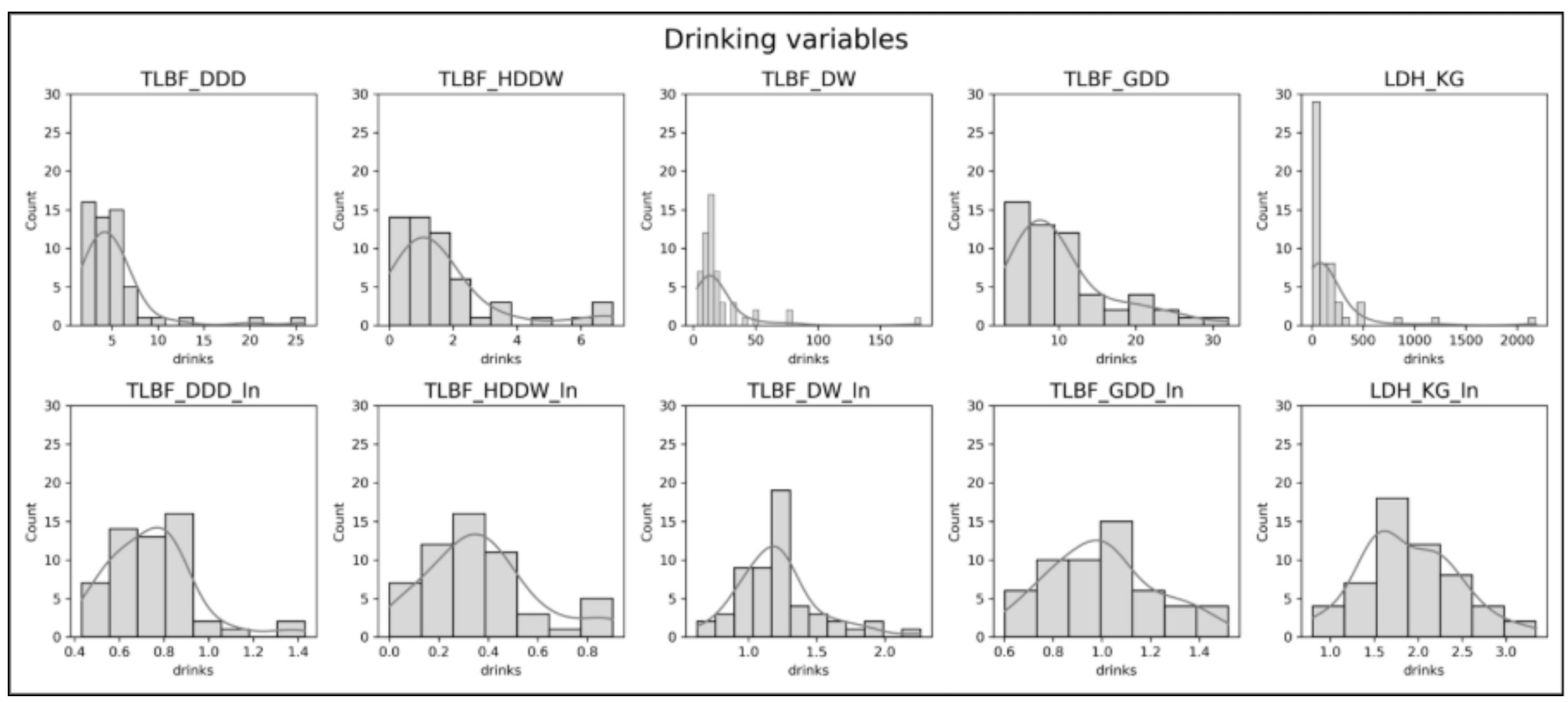

**Figure S1.** Drinking related variables capture recent and long-term drinking patterns. TLFB_DDD: drinks per drinking day, TLFB_HDDW: heavy drinking days per drinking week, TLFB_DW: drinks per week, TLFB_GDD: maximum drinks per drinking day, LDH_KG: lifetime drinking history in kilograms. The original variables presented skewed distributions, so they were logarithmically transformed and used as input to PCA to compute a composite Drinking variable.

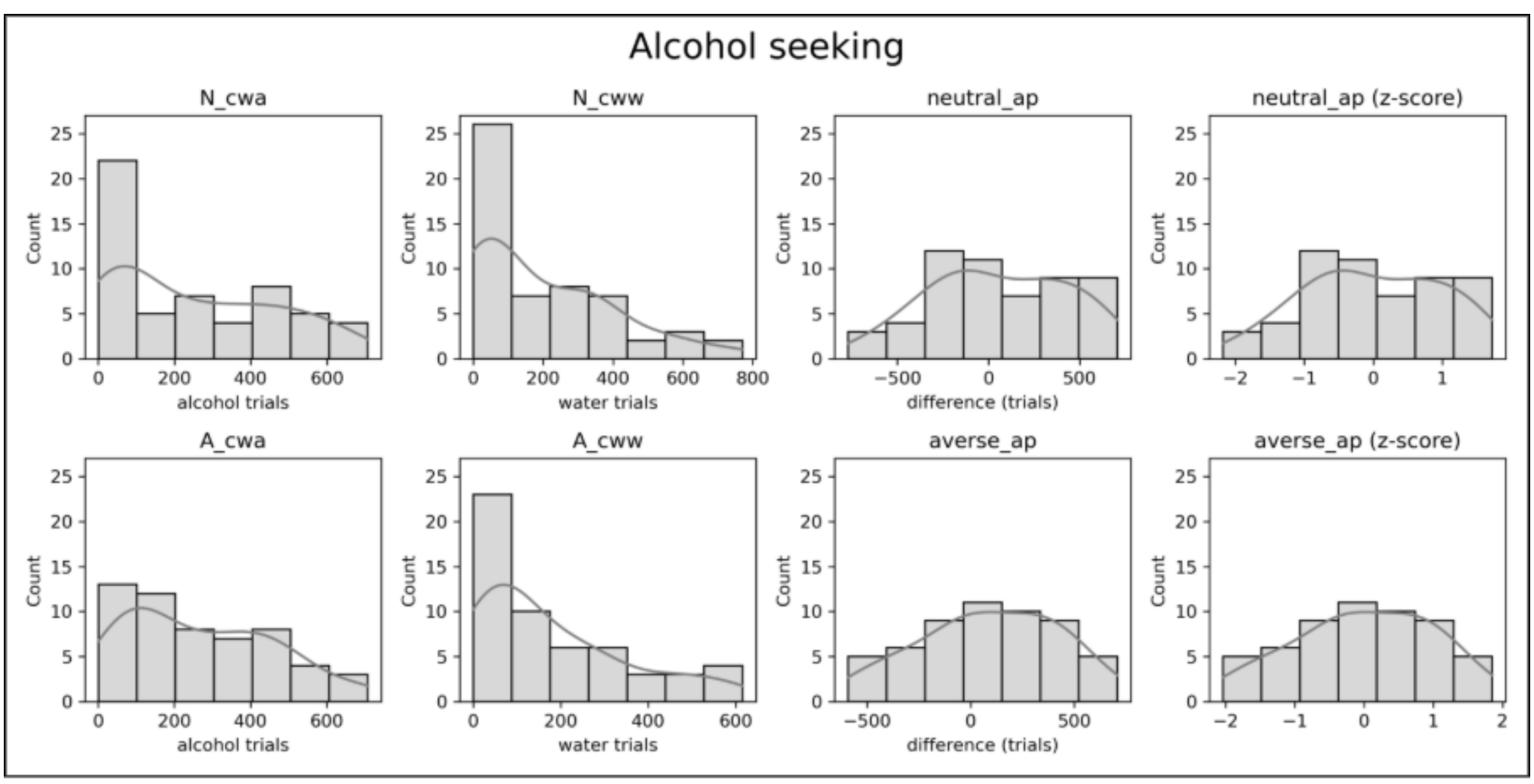

**Figure S2.** Cumulative work (number of trials) for alcohol and water during the alcohol self-administration sessions. A_cwa: cumulative work for alcohol in aversive session. A_cww: cumulative work for water in aversive session. N_cwa: cumulative work for alcohol in neutral session. N_cww: cumulative work for water in neutral session.

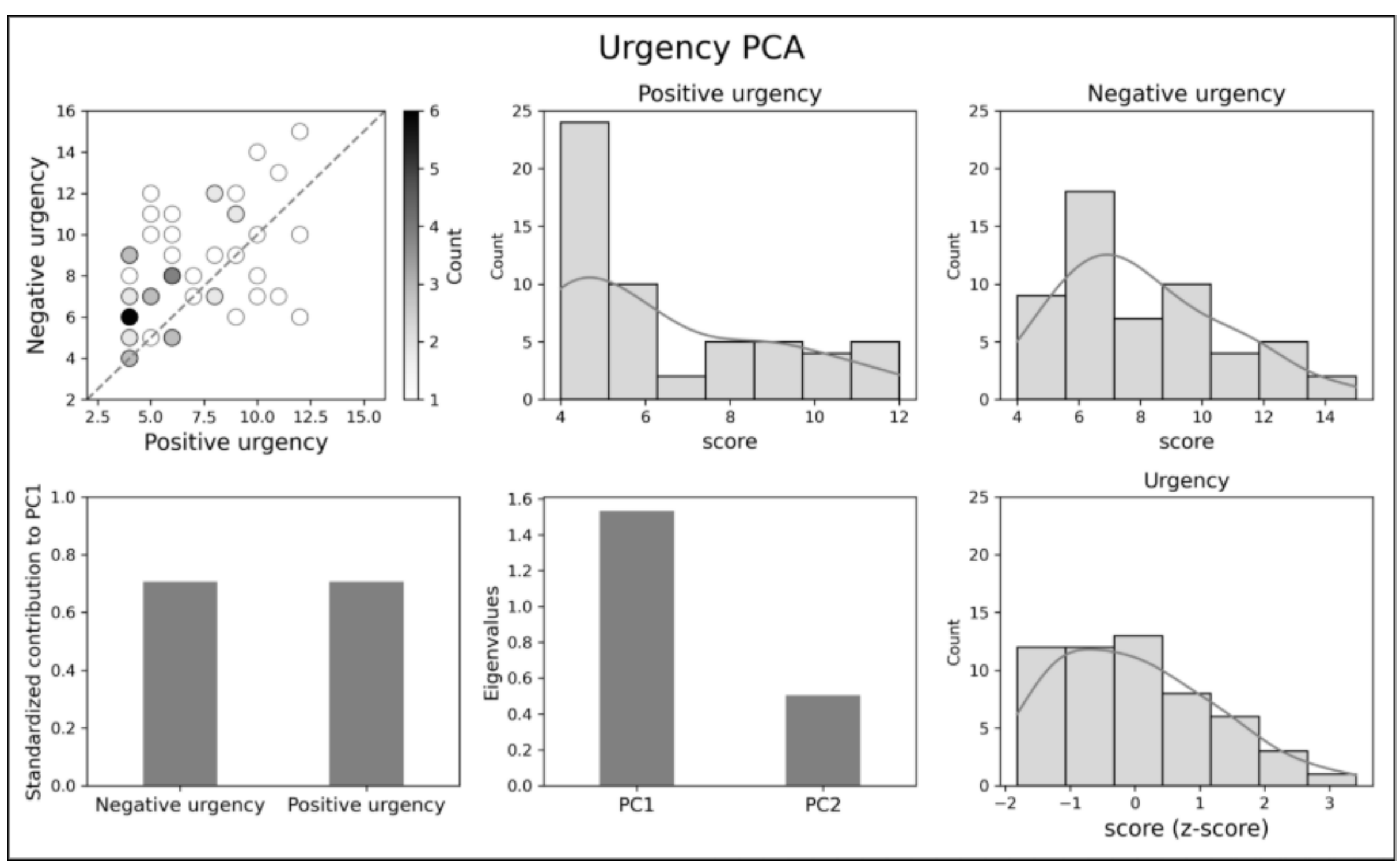

**Figure S3.** Positive and negative urgency are combined into a single urgency variable using PCA (first principal component explained 75% of the variance). The contributions of each individual urgency variable to the first PCA are similar.

Drinking PCA

Correlation between drinking variables (log)

| | TLBF_DDD | TLBF_HDDW | TLBF_DW | TLBF_GDD | LDH_KG |
|---|---|---|---|---|---|
| TLBF_DDD | | | | | |
| TLBF_HDDW | 0.76 | | | | |
| TLBF_DW | 0.75 | 0.81 | | | |
| TLBF_GDD | 0.72 | 0.47 | 0.56 | | |
| LDH_KG | 0.59 | 0.63 | 0.63 | 0.33 | |

Contribution to 1st component

Eigenvalues

PC1 PC2

Drinking

Count

score (z-score)

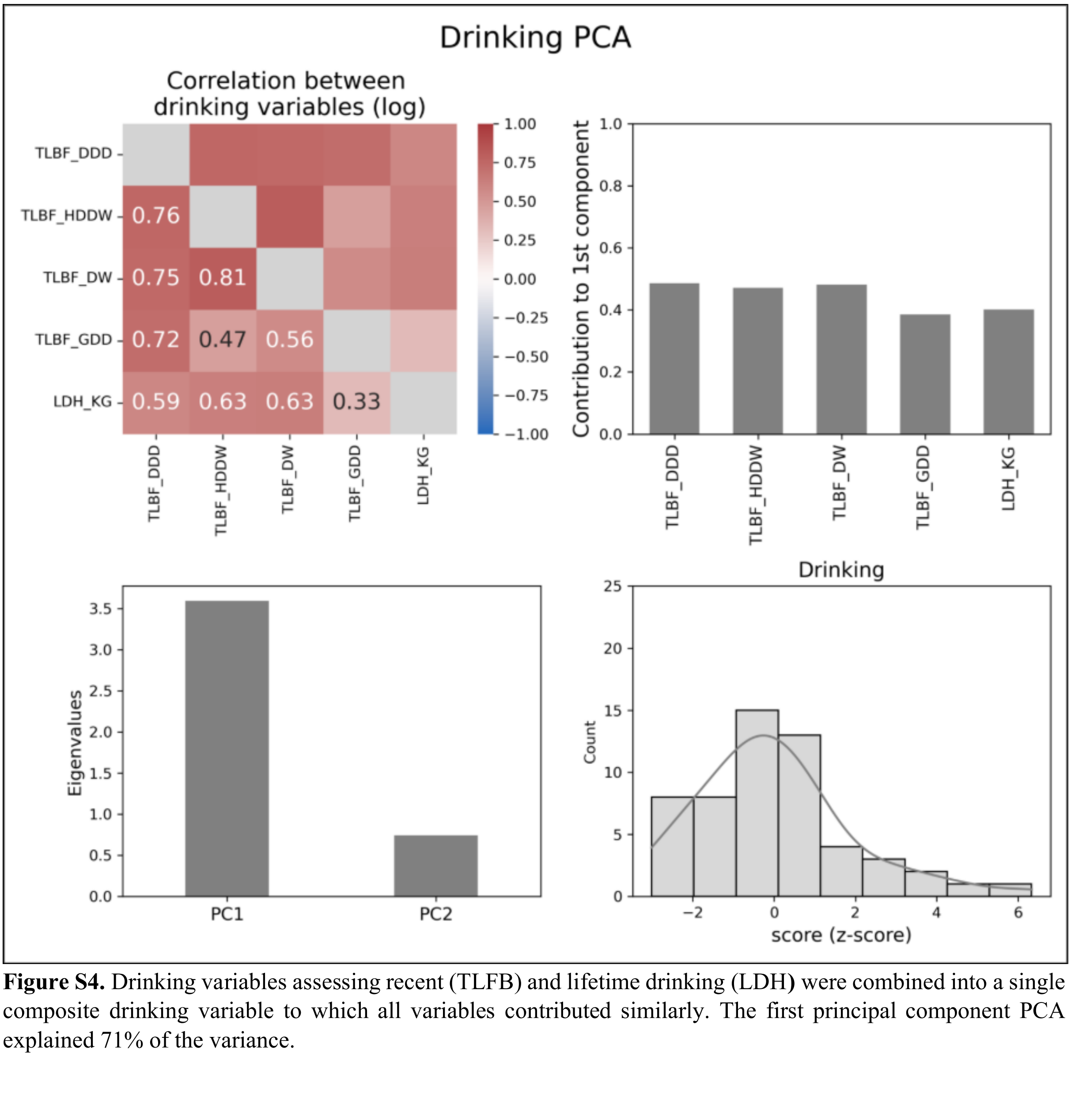

**Figure S4.** Drinking variables assessing recent (TLFB) and lifetime drinking (LDH**)** were combined into a single composite drinking variable to which all variables contributed similarly. The first principal component PCA explained 71% of the variance.

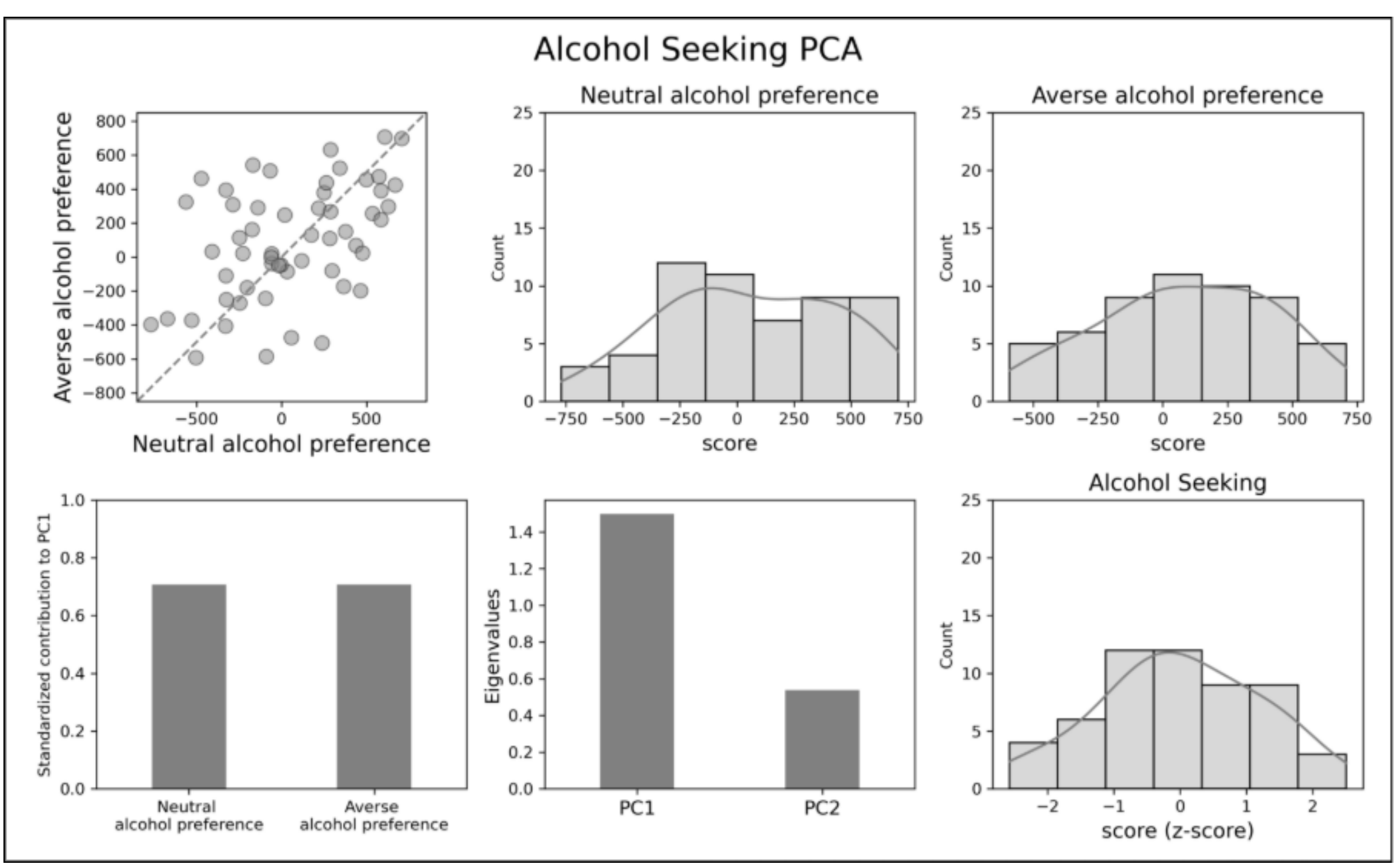

**Figure S5.** Neutral and averse alcohol preferences were combined into a single alcohol seeking variable comprising the general willingness to work for alcohol across both neutral and aversive sessions (first principal component explained 74% of the variance). Contributions of each individual variable to the first PCA are similar.

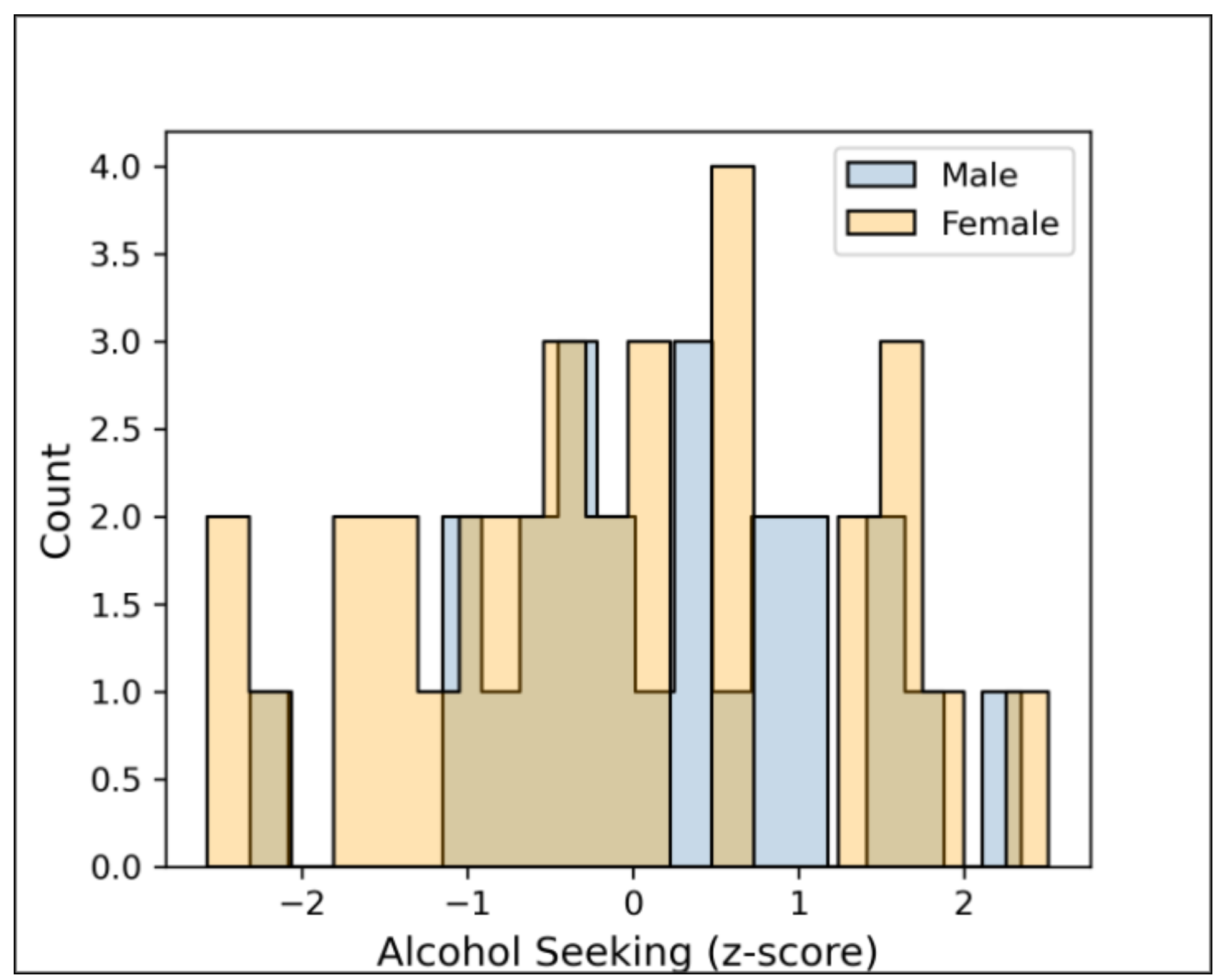

**Figure S6**. Alcohol seeking distributions for male and female participants. There were no significant differences by sex (unpaired t-test, t=1.10 p=0.28).

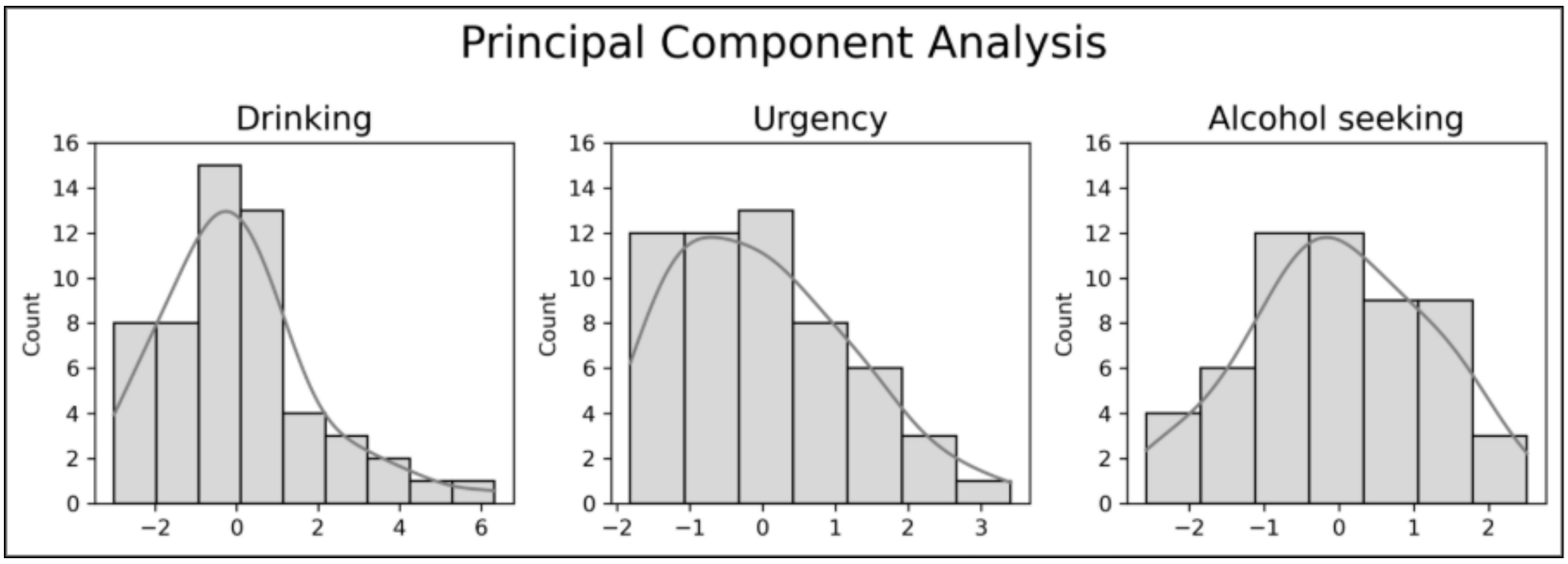

**Figure S7.** The first principal component was obtained for each of the following three sets of variables. **Drinking:** 1st component of TLFB_DDD, TLFB_HDDW, TLFB_DW, TLFB_GDD, LDH_KG (71% explained variance). **Urgency:** 1st component of positive and negative urgency (75% explained variance). **Alcohol seeking:** 1st component of aversive and neutral alcohol preference (74% explained variance).

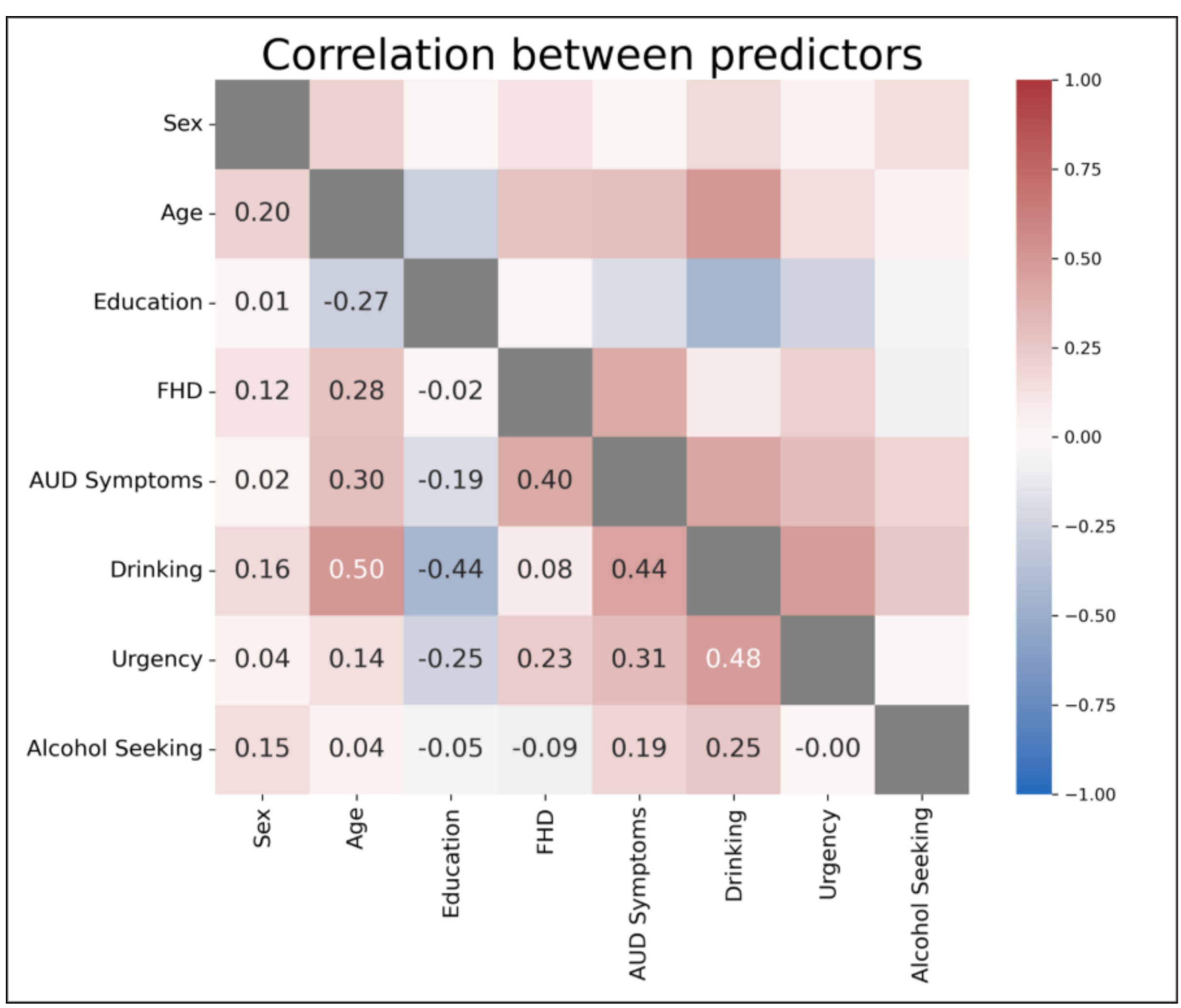

**Figure S8.** Pairwise Pearson's correlation coefficients between the variables spanning the phenotype domain used as input to the Partial Least Square model.

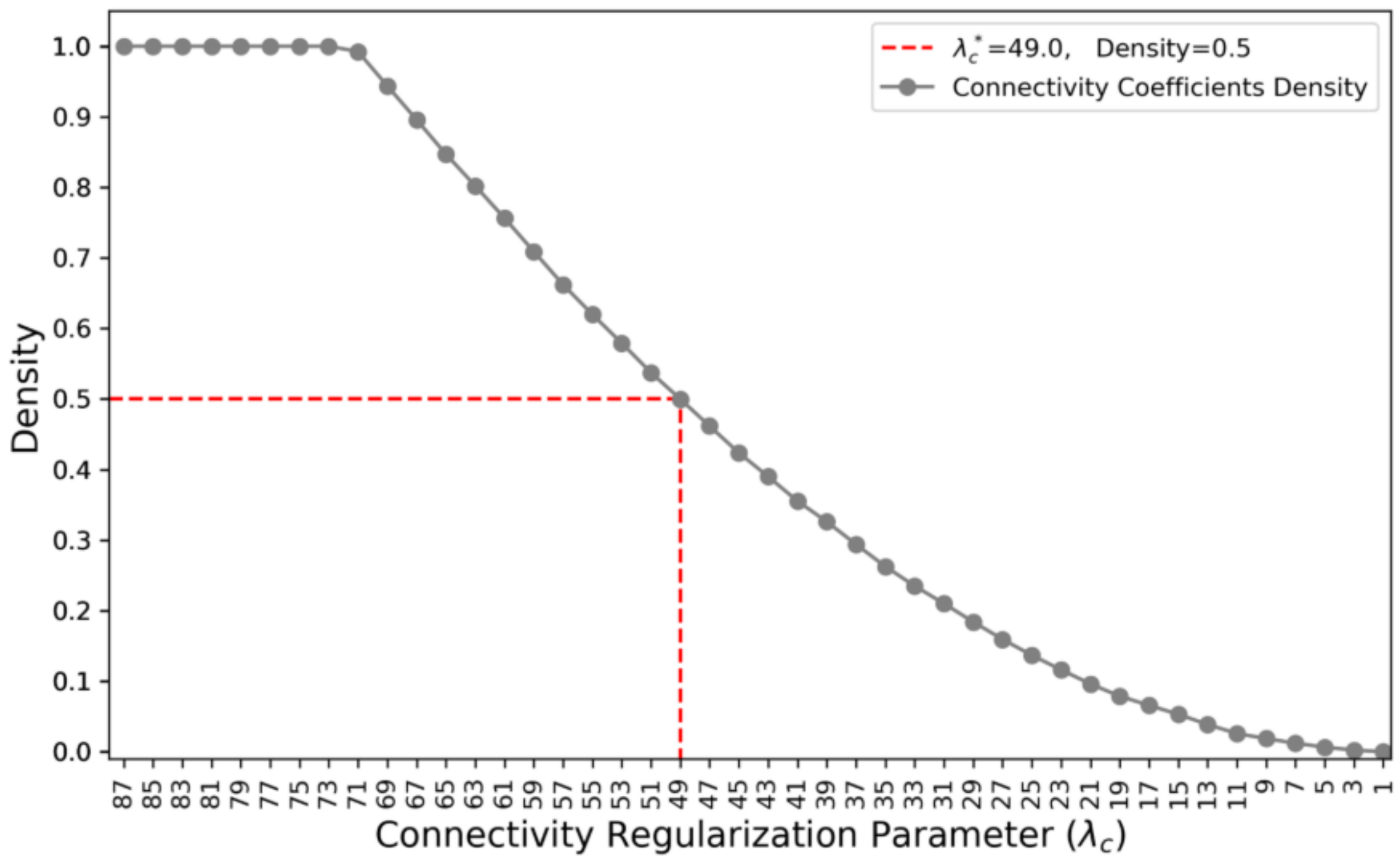

**Figure S9.** The Partial Least Squares regularization parameter (lambda) was set to the value 49.0. At this value, the density of the connectivity solution (number of functional edges participating in the solution) is 50%.

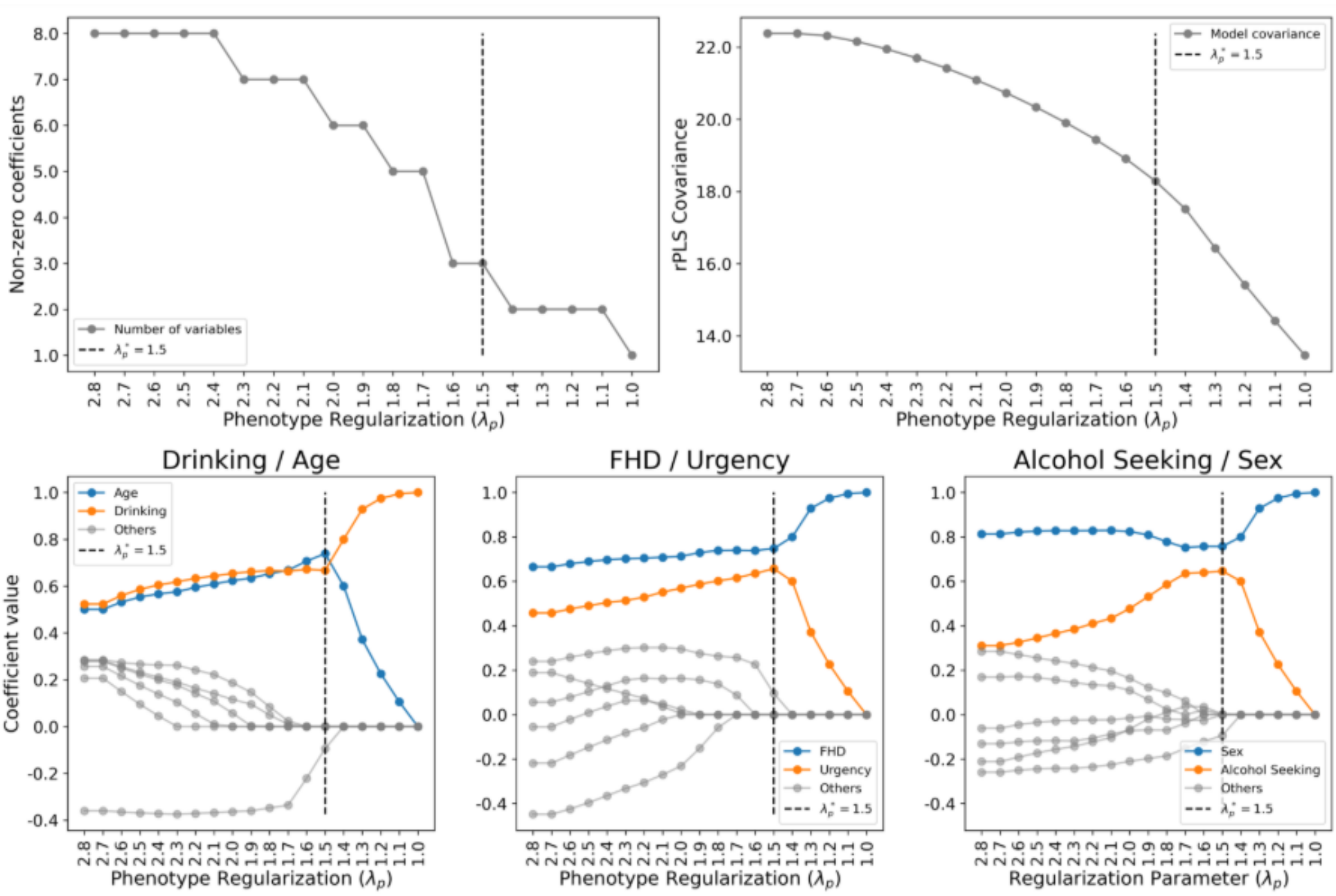

**Figure S10**. The regularization parameter for the phenotypic domain (1.5 indicated by the gray horizontal line) was selected based on the number of non-zero coefficients in component 1.

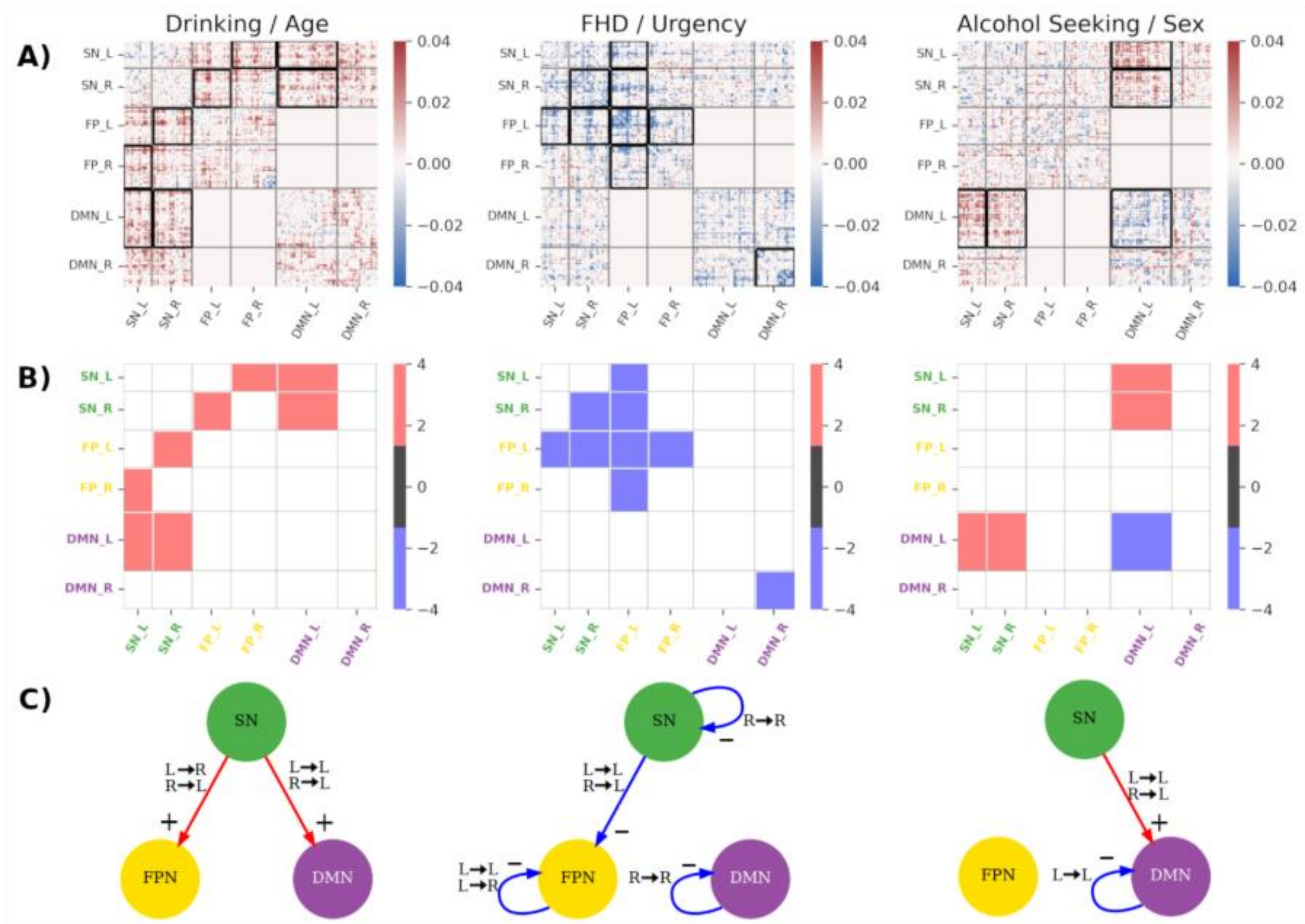

**Figure S11.** Statistical significance of within and between lateralized network interactions for each rPLS component. The statistical significance of the interactions was assessed using a null model distribution for interaction strength. Significant network interactions are shown as block contours (A), and the sign of the interaction (B) is denoted by red (positive) and blue (negative). Resulting diagrams incorporating directionality from the TNM are shown in (C).

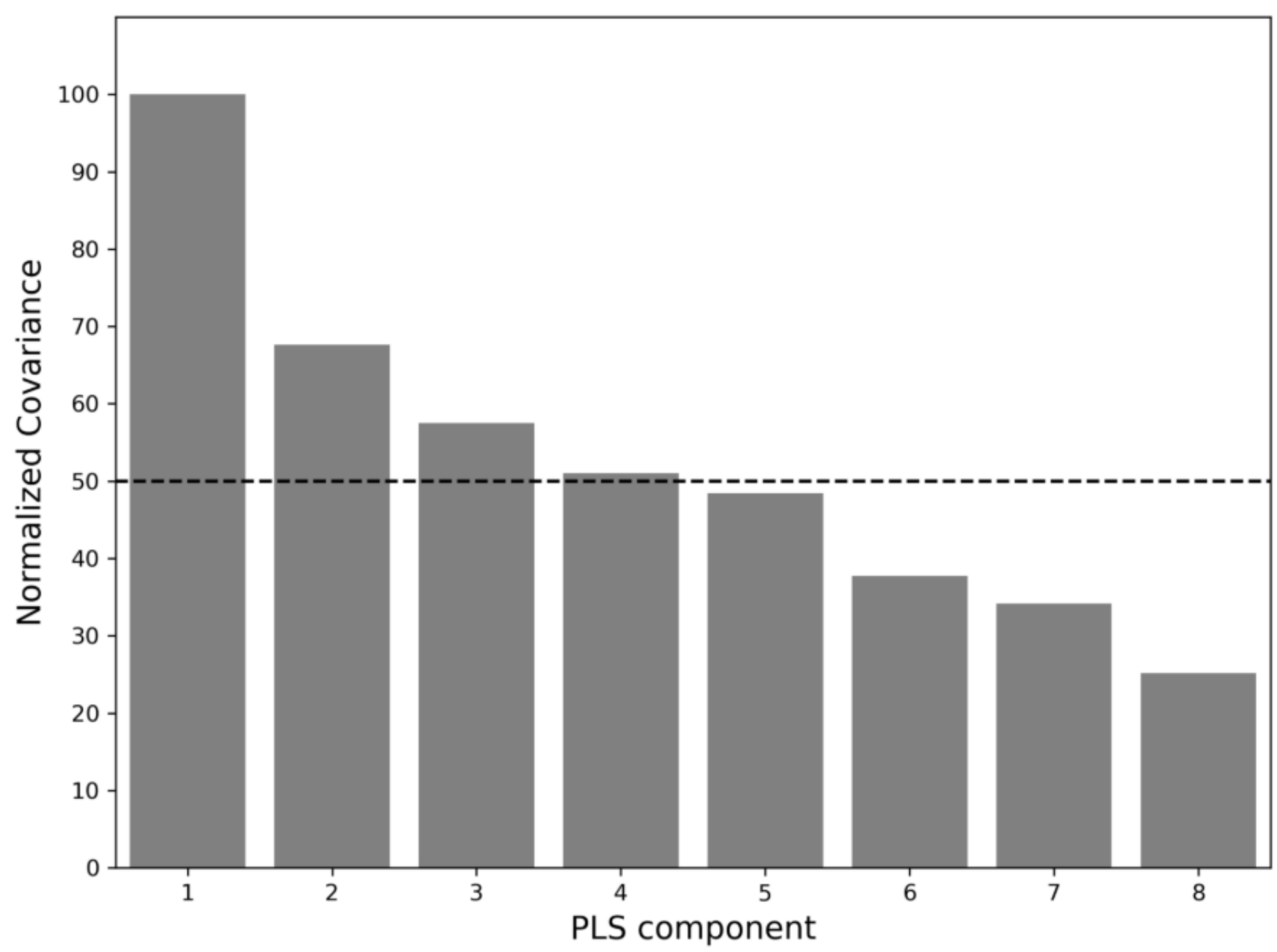

**Figure S12**. Covariance associated with each of the eight PLS components, normalized with respect to the maximum covariance (Component 1: 100%). A threshold criterion of 50% for the relative covariance retained the first four PLS components for subsequent analyses.

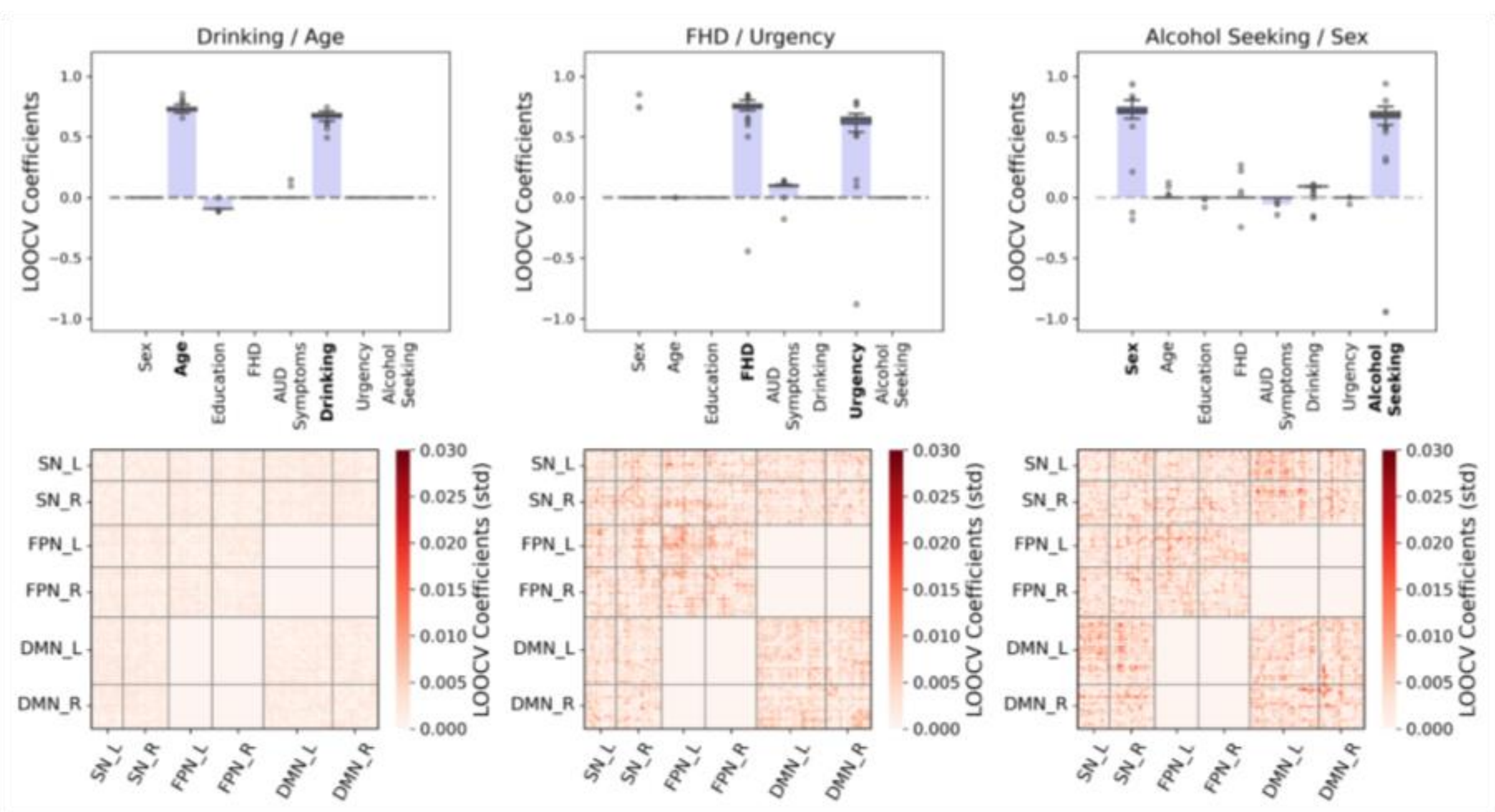

**Figure S13.** Leave-one-out cross validation of rPLS results (LOOCV; N=55) for the regularized PLS components. Top row: variability of the phenotypic characteristic coefficients across the LOOCV runs (boxplots) and full cohort coefficients for reference (purple bars). Bottom row: variability (standard deviation) of coefficients associated with each functional coupling along the LOOCV runs for each component. Interactions between FPN and DMN are excluded from the analysis following the assumptions of the TNM.

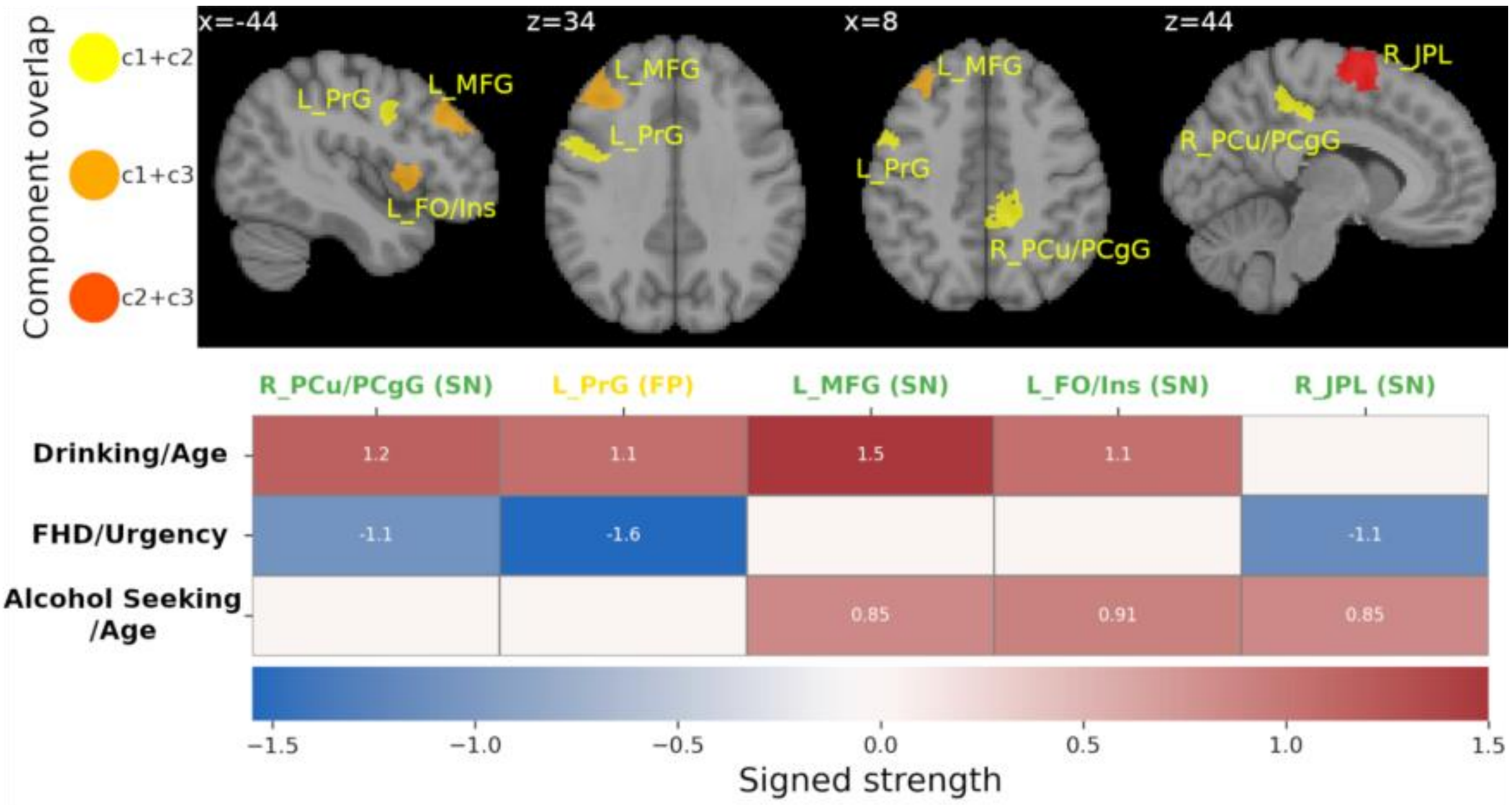

**Figure S14.** Overlapping regions between rPLS components. Among the top contributing regions in each alcohol-related component, five regions were found to participate in more than one component: the right precuneus/posterior cingulate gyrus (PCu/PCgG), left precentral gyrus (PrG), left middle frontal gyrus (MFG), left frontal operculum/insula (FO/Ins), and right juxtapositional lobule cortex (JPL). SN: Salience Network, FP: Frontoparietal Network.

## Drinking/Age Component

| | Network | Hemisphere | Schaefer 300 index | Schaefer 300 region | Description | Sign | Strength |
|---|---|---|---|---|---|---|---|
| 1 | SN | R | 214 | SalVentAttnA_ParOper_4 | SMG | + | 1.73 |
| 2 | FPN | R | 248 | ContA_IPS_3 | SPL/SMG | + | 1.58 |
| 3 | SN | R | 229 | SalVentAttnB_PFCl_2 | FP | + | 1.54 |
| 4 | SN | R | 212 | SalVentAttnA_ParOper_2 | PO | + | 1.52 |
| 5 | SN | L | 76 | SalVentAttnB_PFCl_2 | MFG | + | 1.50 |
| 6 | SN | L | 73 | SalVentAttnA_FrMed_2 | SFG | + | 1.29 |
| 7 | SN | R | 217 | SalVentAttnA_Ins_2 | pINS | + | 1.27 |
| 8 | SN | L | 68 | SalVentAttnA_Ins_3 | CO/mINS | + | 1.25 |
| 9 | SN | L | 72 | SalVentAttnA_FrMed_1 | ACgG/JPL | + | 1.20 |
| 10 | SN | R | 222 | SalVentAttnA_ParMed_2 | PCu/PCgG | + | 1.17 |
| 11 | DMN | L | 113 | DefaultA_PFCd_2 | SFG/MFG | + | 1.13 |
| 12 | SN | L | 75 | SalVentAttnB_PFCl_1 | FP | + | 1.10 |
| 13 | FPN | L | 99 | ContA_PFCl_4 | PrG | + | 1.06 |
| 14 | DMN | R | 284 | DefaultB_AntTemp_1 | R-TP | + | 1.05 |
| 15 | SN | L | 75 | SalVentAttnB_Ins_2 | FO/Ins | + | 1.05 |

## Family History/Urgency Component

| | Network | Hemisphere | Schaefer 300 index | Schaefer 300 region | Description | Sign | Strength |
|---|---|---|---|---|---|---|---|
| 1 | SN | R | 223 | SalVentAttnA_ParMed_3 | PoG/PrG | - | 1.59 |
| 2 | FPN | L | 99 | ContA_PFCl_4 | PrG | - | 1.55 |
| 3 | SN | R | 215 | SalVentAttnA_PrC_1 | PrG | - | 1.31 |
| 4 | FPN | L | 100 | ContA_Cingm_1 | ACgG | - | 1.30 |
| 5 | SN | L | 71 | SalVentAttnA_ParMed_2 | PCu | - | 1.28 |
| 6 | DMN | R | 288 | DefaultB_PFCv_2 | IFG | - | 1.23 |
| 7 | DMN | R | 289 | DefaultB_PFCv_3 | IFG | - | 1.18 |
| 8 | SN | R | 224 | SalVentAttnA_FrMed_2 | JPL | - | 1.15 |
| 9 | SN | L | 70 | SalVentAttnA_ParMed_1 | PCu/PCgG | - | 1.14 |
| 10 | FPN | L | 110 | ContC_Cingp_1 | PCgG | - | 1.13 |
| 11 | SN | R | 221 | SalVentAttnA_ParMed_1 | pCgG/aCgG | - | 1.11 |
| 12 | FPN | R | 247 | ContA_IPS_2 | SMG | - | 1.08 |
| 13 | SN | R | 222 | SalVentAttnA_ParMed_2 | PCu/PCgG | - | 1.08 |
| 14 | SN | R | 233 | SalVentAttnB_PFCmp_2 | SFG | - | 1.07 |
| 15 | FPN | L | 91 | ContA_IPS_2 | L-SPL/SMG | - | 1.07 |

## Alcohol Seeking/Sex Component

| | Network | Hemisphere | Schaefer 300 index | Schaefer 300 region | Description | Sign | Strength |
|---|---|---|---|---|---|---|---|
| 1 | DMN | R | 277 | DefaultA_pCunPCC_3 | PCu | + | 1.54 |
| 2 | SN | L | 67 | SalVentAttnA_Ins_2 | FO | + | 1.16 |

| 3 | SN | R | 216 | SalVentAttnA_Ins_1 | aINS | + | 1.13 |
|---|---|---|---|---|---|---|---|
| 4 | SN | L | 65 | SalVentAttnA_ParOper_2 | PO/SMG | + | 0.91 |
| 5 | SN | R | 220 | SalVentAttnA_FrMed_1 | R-ACgG | + | 0.85 |
| 6 | SN | L | 78 | SalVentAttnB_Ins_2 | FO/Ins | + | 0.81 |
| 7 | DMN | L | 126 | DefaultB_IPL_1 | AG | + | 0.74 |
| 8 | FPN | R | 251 | ContA_PFCl_2 | IFG/MFG | + | 0.71 |
| 9 | DMN | L | 114 | DefaultA_pCunPCC_1 | PCu | + | 0.71 |
| 10 | SN | R | 224 | SalVentAttnA_FrMed_2 | JPL | + | 0.70 |
| 11 | SN | L | 76 | SalVentAttnB_PFCl_2 | MFG | + | 0.68 |
| 12 | SN | L | 69 | SalVentAttnA_FrOper_1 | PCu/PCgG | - | 0.99 |
| 13 | SN | L | 74 | SalVentAttnB_IPL_1 | SMG | - | 0.79 |
| 14 | SN | L | 64 | SalVentAttnA_ParOper_1 | SMG | - | 0.68 |
| 15 | DMN | R | 282 | DefaultA_PFCm_5 | ACgG | - | 0.66 |

Education Component

| | Network | Hemisphere | Schaefer 300 index | Schaefer 300 region | Description | Sign | Strength |
|---|---|---|---|---|---|---|---|
| 1 | SN | R | 221 | SalVentAttnA_ParMed_1 | pCgG/aCgG | + | 1.22 |
| 2 | DMN | R | 289 | DefaultB_PFCv_3 | IFG | + | 1.12 |
| 3 | DMN | L | 141 | DefaultC_Rsp_1 | pCgG | + | 1.01 |
| 4 | DMN | L | 112 | DefaultA_PFCd_1 | SFG | + | 1.00 |
| 5 | DMN | L | 132 | DefaultB_PFCd_5 | SFG | + | 0.92 |
| 6 | SN | R | 222 | SalVentAttnA_ParMed_2 | PCu/PCgG | + | 0.92 |
| 7 | DMN | R | 274 | DefaultA_PFCd_1 | MFG | + | 0.91 |
| 8 | SN | R | 226 | SalVentAttnB_IPL_1 | FP | - | 1.45 |
| 9 | SN | R | 219 | SalVentAttnA_FrOper_1 | CO | - | 1.41 |
| 10 | SN | L | 73 | SalVentAttnA_FrMed_2 | SFG | - | 1.31 |
| 11 | SN | R | 216 | SalVentAttnA_Ins_1 | aINS | - | 1.17 |
| 12 | SN | R | 218 | SalVentAttnA_Ins_3 | CO/mINS | - | 0.91 |
| 13 | SN | R | 221 | SalVentAttnA_ParMed_1 | pCgG/aCgG | - | 0.91 |
| 14 | SN | R | 217 | SalVentAttnA_Ins_2 | pINS | - | 0.90 |
| 15 | SN | R | 224 | SalVentAttnA_FrMed_2 | JPL | - | 0.84 |

**Table S1.** Top brain regions per component (top 5%, 15 regions). Index refers to Schaefer 300 parcellation (Schaefer et al., 2018) ACgG: Anterior Cingulate Gyrus, ACgG/JPL: Anterior Cingulate Gyrus/Juxtapositional Lobule, aINS: Ventral Anterior Insula, aINS/OFC: Anterior Insula/Orbitofrontal Cortex, CO/mINS: Central Operculum/Middle Insular Cortex, FO: Frontal Operculum, FO/OFC: Frontal Operculum/Lateral Orbitofrontal Cortex, FrP: Frontal Pole, IFG: Inferior Frontal Gyrus, IFG/MFG: Inferior/Middle Frontal Gyrus, INS: Insula, JPL: Juxtapositional Lobule Cortex, LOC: Lateral Occipital Cortex (superior), MFG: Middle Frontal Gyrus, MFG/IFG: Middle/Inferior Frontal Gyrus, MTG: Middle Temporal Gyrus, PCgG: Posterior Cingulate Cortex (retrosplenial), pCgG/aCgG: Cingulate Gyrus (posterior and anterior), PCu: Precuneus (anterior/dorsal), PCu/PCgG: Precuneus/Posterior Cingulate Gyrus, pINS: Ventral Posterior Insula, PO: Parietal Operculum, PoG/PrG: Postcentral/Precentral Gyrus (medial), extends into posterior Cingulate Gyrus (anteriorly) and Precuenus (posteriorly), PrG: Precentral Gyrus, PrG/IFG: Precentral/Inferior Frontal Gyrus, SFG: Superior Frontal Gyrus/Paracingulate Gyrus, SFG/MFG: Superior/Middle Frontal Gyrus, SMG: Supramarginal Gyrus (anterior), SPL/SMG: Superior Parietal Lobule/Supramarginal Gyrus (posterior)